\documentclass[12pt]{article}
\usepackage{comment}
\usepackage{textcomp}
\usepackage{chetnew}
\usepackage{amssymb,amsmath,amsfonts,mathrsfs,braket}
\usepackage[utf8]{inputenc}
\usepackage{tikz,tikz-cd,url}
\usepackage{xcolor}
\usetikzlibrary{arrows,decorations.markings,shapes.geometric,decorations.pathmorphing}
\tikzset{snake it/.style={decorate, decoration=snake}}

\usepackage{bm}

\renewcommand{\d}[1]{\ensuremath{\operatorname{d}\!{#1}}}

\def\one{{\,\hbox{1\kern-.8mm l}}}

\newcommand{\wn}{\widehat{\nabla}}
\newcommand{\ws}{\widehat{\square}}
\newcommand{\m}{\lambda}

\newcommand{\pd}{\partial}
\newcommand{\rank}{\text{rank}}

\newcommand{\Tr}{\mathrm{Tr}}

\allowdisplaybreaks

\def\makeatletter{\catcode`\@=11}
\makeatletter
\def\mathbox#1{\hbox{$\m@th#1$}}%
\def\math@ccstyles#1#2#3#4#5#6#7{{\leavevmode
      \setbox0\mathbox{#6#7}%
      \setbox2\mathbox{#4#5}%
      \dimen@ #3%
      \baselineskip\z@\lineskiplimit#1\lineskip\z@
      \vbox{\ialign{##\crcr
             \hfil \kern #2\box2 \hfil\crcr
             \noalign{\kern\dimen@}%
             \hfil\box0\hfil\crcr}}}}
\def\mathaccstyles{\math@ccstyles\maxdimen}
\def\maththroughstyles{\math@ccstyles{-\maxdimen}}
\def\unity%
 {\maththroughstyles{.45\ht0}\z@\displaystyle {\mathchar"006C}\displaystyle 1}

\def\AA{{\cal A}}

\def\II{{\cal I}}
\def\JJ{{\cal J}}

\def\LL{{\cal L}}
\def\MM{{\cal M}}
\def\NN{{\cal N}}
\def\OO{{\cal O}}

\def\TT{{\cal T}}

\def\VV{{\cal V}}

\def\Tr{{\rm {Tr}}}

\def\d{{\partial}}

\def\beq{\begin{equation}}
\def\eeq{\end{equation}}
\newcommand{\bea}{\begin{eqnarray}}
\newcommand{\eea}{\end{eqnarray}}
\def\bal{\begin{align}}
\def\eal{\end{align}}

\preprint{CCTP-22-6 \hfill  \\ ITCP-IPP-2022/7\hfill
  DESY-21-169}

\title{\vspace{-1.cm} Covariantly Constant Anomalies\\ \vspace{0.4cm} 
on Conformal Manifolds
}

\author{
Enrico Andriolo$^{a,\clubsuit}$,
Vasilis Niarchos\;$^{b,\diamondsuit}$, Constantinos Papageorgakis\;$^{a,\spadesuit}$, Elli Pomoni\;$^{c,\heartsuit}$}

\affiliation{
$^a$ Centre for Theoretical Physics, Department of Physics and Astronomy\\ Queen Mary University of London, London E1 4NS, UK \vspace{0.3cm} $ $ \\
$^b$ ITCP \& CCTP, Department of Physics,\\
University of Crete, 71003 Heraklion, Greece
\vspace{0.3cm} $ $\\

$^c$DESY, Theory Group, Notkestrasse 85, Building 2a, 22607 Hamburg, Germany \\

\vspace{0.3cm}
{\tt \small$^\clubsuit$e.andriolo@qmul.ac.uk,
$^\diamondsuit$niarchos@physics.uoc.gr,  
$^\spadesuit$c.papageorgakis@qmul.ac.uk, 

$^{\heartsuit}$ elli.pomoni@desy.de}}

\abstract{Operators with integer scaling dimensions in even-dimensional conformal field theories exhibit well-known type-B Weyl anomalies. In general, these anomalies depend non-trivially on exactly marginal couplings. We study the corresponding fully covariantised anomaly functional on conformal manifolds in several examples. We show that a natural consequence of the Wess--Zumino consistency condition is that the anomalies are covariantly constant with respect to the exactly marginal couplings. The argument is general and applies even when the conformal symmetry is spontaneously broken on moduli spaces of vacua.  
}

\date{}

\begin{document}

\maketitle

\hypersetup{pageanchor=true}

\setcounter{tocdepth}{2}

\toc

\section{Introduction and Summary of Results}\label{summary}

The fate of Weyl anomalies under deformations is an interesting subject in its own right. One can consider deformations that preserve the conformal symmetry, or break it explicitly or spontaneously. In principle, these properties can be used to constrain non-perturbative Quantum Field Theory (QFT) dynamics. For a recent review see \cite{Niarchos:2022ljh}.

In this paper, we will focus exclusively on four-dimensional CFTs. The question of whether conformal anomalies match in different phases of a 4D CFT was answered in the affirmative in \cite{Schwimmer:2010za} for type-A anomalies, and put to great use in the proof of the a-theorem \cite{Komargodski:2011vj}. In general, it is not possible to obtain similar results for type-B Weyl anomalies.

The non-perturbative properties of type-B Weyl anomalies associated with Coulomb-branch Operators (CBOs) on the Higgs--branch vacuum moduli space of 4D $\NN=2$ superconformal CFTs (SCFTs) were discussed in \cite{Niarchos:2019onf,Niarchos:2020nxk}. These papers presented examples where the CBO type-B Weyl anomalies matched across the Higgs branch.\footnote{In previous versions of \cite{Niarchos:2019onf,Niarchos:2020nxk} there was a reported mismatch between the anomaly in the conformally symmetric and spontaneously broken phases leading to a number of non-perturbative conjectures, which were postulated in \cite{Niarchos:2020nxk}. This mismatch has been resolved in the latest versions of the papers, in agreement with the general arguments of \cite{Schwimmer:2023nzk}.}

In the present work, we elaborate further on the properties of the CBO type-B Weyl anomalies, and point out that one of the crucial elements in the discussion of Refs\ \cite{Niarchos:2019onf,Niarchos:2020nxk}---the fact that these anomalies are covariantly constant on conformal manifolds---can be understood in many cases as a natural consequence of the Wess--Zumino consistency conditions of the corresponding anomaly functionals. This alternative perspective is useful for reasons that we will explain. The existence of covariantly constant type-B anomalies in different phases of the theory has non-trivial implications as explained in Ref.\ \cite{Niarchos:2020nxk} and reviewed in \cite{Niarchos:2022ljh}.

The main elements of the argument are as follows. For an operator $\OO$ with scaling dimension $\Delta=2+n$ $(n\in \mathbb N_0)$, the anomaly of interest can be identified (in all phases of the CFT) as a specific contact term in the integrated 3-point function of the trace of the energy-momentum tensor $T\equiv {T^\mu}_\mu$,
\beq
\label{sumaa}
\int d^4 y \langle T(y) \OO_\Delta(x) \OO_\Delta(0)\rangle \propto \Box^n \delta(x)
~.
\eeq
In the unbroken conformal phase, the Ward identities of diffeomorphism and Weyl transformations can be used to relate the corresponding anomaly coefficient $G_\Delta^{(\rm CFT)}$, to the 2-point function coefficient of the operator $\OO$. In momentum space, the anomaly appears in the logarithmically divergent piece of the 2-point function
 \begin{equation}
\label{extractingCFT}
\langle \OO_{\Delta}(p) \OO_{\Delta}(-p)\rangle =(-1)^{n+1} \frac{\pi^2 G_\Delta^{(\rm CFT)}}{2^{2n} \Gamma(n+1)\Gamma(n+\frac{D}{2})} p^{2n} \log \left( \frac{p^2}{\mu^2}\right)+...
~.
\end{equation}
In a phase with spontaneous breaking of the conformal symmetry, the Ward identities do not provide a similar relation between the corresponding type-B conformal anomaly and some datum in the 2-point function of the operator $\OO$. In that case, the broken-phase anomaly, $G_{\Delta}^{(\rm broken)}$, must be extracted directly from the 3-point function \eqref{sumaa}. In momentum space this reads
\begin{align}
\label{extractingHB}
\lim_{q\to0}\langle T (q) \OO_\Delta(p_1) \OO_\Delta(p_2) \rangle &= (-1)^n\frac{\pi^2 G_{\Delta}^{(\rm broken)} }{2^{2n} \Gamma(n+1)\Gamma(n+2)} (p_1^2)^n+... ~.
\end{align}
On the RHS of \eqref{extractingCFT} and \eqref{extractingHB} a Dirac-delta imposing momentum conservation is left implicit; in particular, the $q\to0$ limit in \eqref{extractingHB} is equivalent to taking the $p_2\to-p_1$ limit. 

As an explicit example, let us consider the case of 4D $\NN=2$ SCFTs with a non-trivial chiral ring of CBOs $\OO_I$, an anti-chiral ring of conjugate operators $\bar \OO_J$, and a non-empty conformal manifold $\MM$. The latter means that the $\NN=2$ SCFTs of interest possess exactly marginal operators.\footnote{These are necessarily supersymmetric descendants of scaling-dimension 2 CBOs.} The CBOs, which are charged by default under the $U(1)_r$ part of the full $U(1)_r \times SU(2)_R$ R-symmetry of the theory, have integer scaling dimensions, and the corresponding type-B Weyl anomalies can be obtained as contact terms in the $U(1)_r$-preserving 3-point functions $\langle T(y) \OO_I(x) \bar \OO_J(0)\rangle$. In the conformally symmetric phase we will denote the corresponding anomaly coefficients as $G_{I\bar J}^{(\rm CFT)}$. On the Higgs branch\footnote{4D $\NN=2$ SCFTs typically have both Higgs and Coulomb branch moduli spaces of vacua. Here we consider only the case of Higgs moduli spaces to make contact with the discussion in Refs.\ \cite{Niarchos:2019onf,Niarchos:2020nxk}.} the $SU(2)_R$ part of the R-symmetry is spontaneously broken along with conformal symmetry by the non-vanishing vacuum expectation values (vevs) of Higgs--branch superconformal primary operators. We will denote the corresponding anomaly coefficients in this phase $G_{I\bar J}^{(\rm Higgs)}$. In this context, we are mainly interested in the properties of $G_{I\bar J}^{(\rm CFT)}$ and $G_{I\bar J}^{(\rm Higgs)}$, but we will soon indicate which arguments of the paper can be generalised beyond these specific cases. Both quantities are, in general, complicated functions of the exactly marginal couplings (see \cite{Niarchos:2019onf,Niarchos:2020nxk} and references therein).

A crucial ingredient in the discussion of \cite{Niarchos:2019onf,Niarchos:2020nxk} was the proposal that the anomaly coefficients $G_{I\bar J}$ are covariantly constant on the conformal manifold $\mathcal M$ in both phases of the theory. Namely, both anomalies obey equations of the form $\nabla G_{I\bar J}^{(\rm CFT)}=0$, $\nabla G_{I\bar J}^{(\rm Higgs)}=0$, where $\nabla$ is a phase-independent connection on the vector bundles of the CBOs. It is straightforward to derive this condition in the conformally symmetric phase as a consequence of superconformal Ward identities. However, as pointed out in \cite{Niarchos:2019onf}, a similar argument in the Higgs phase needs to take into account potential contributions from the dilatino. In \cite{Niarchos:2019onf} it was anticipated that such contributions do not affect the contact term that accounts for the anomaly, but it is not straightforward to demonstrate this explicitly. As a result, it would be very useful to have an independent argument that $\nabla G_{I\bar J}^{(\rm Higgs)}=0$. Our main purpose in this note is to find such an argument. As a bonus, the argument we will present is very general and applies to any CFT with a conformal manifold that has operators with integer-valued scaling dimension; it is not restricted to CBOs in $\NN=2$ SCFTs or to Higgs-branch phases.

It is well known (see e.g.\ \cite{Osborn:1991gm}) that conformal anomalies can be conveniently packaged into a {\it local} anomaly functional that expresses the Weyl variation of the generating functional of correlation functions $W$\footnote{See also \cite{Jack:1990eb,Jack:2013sha}.}
\begin{align}
  \label{eq:27}
  \delta_\sigma W \propto \int d^4x \sqrt \gamma \delta\sigma \mathcal A\;.
\end{align}
$W$ is a {\it non-local} functional of the sources of the CFT, but the Weyl anomaly $\mathcal A$ is a local term reflecting the above-mentioned fact that in correlation functions it appears as a contact term. The $\delta_\sigma$ variation in \eqref{eq:27} denotes infinitesimal local Weyl transformations with parameter $\delta\sigma(x)$ that vanish at the boundary of spacetime \cite{Schwimmer:2010za}, and $\gamma_{\mu\nu}$ is the background spacetime metric. The locality of $\delta\sigma(x)$ guarantees, among other things, that the Ward identities retain the same form in all phases of the theory, irrespective of whether or not conformal symmetry is spontaneously broken (they are operatorial relations). This fact will be crucial for our upcoming discussion of the structure of the anomaly functional in different phases. In order to encode the CBO type-B anomalies of interest in the anomaly functional one needs to add to the action spacetime-dependent sources for the operators $\OO_I, \bar \OO_J$
\begin{align}
  \label{eq:28}
  \delta S=\int d^{4} x\sqrt \gamma\left(\lambda^I(x) \mathcal{O}_I(x)+\bar{\lambda}^J(x) \bar{\mathcal{O}}_J(x)\right)\;.
\end{align}

The anomaly functional must satisfy certain conditions. It must be invariant under diffeomorphisms or any other unbroken symmetries of the theory. In addition, it must obey the Wess--Zumino (WZ) consistency condition 
\begin{align}
  \label{eq:11}
  \delta_{[\sigma_{1}} \delta_{\sigma_{2}]} W=0
  ~,
\end{align}
which encodes the fact that the action of the Weyl group is abelian. Finally, terms in $\AA$ that are Weyl variations of a {\it local} functional express the addition of local counterterms in $W$ \cite{Bonora:1985cq}, which simply correspond to a change in the regularisation scheme. Such terms are considered trivial and can be dropped from $\delta_\sigma W$. This reflects the fact that the anomaly is a scheme-independent quantity.

As emphasised already in Ref.\ \cite{Gomis:2015yaa}, on a conformal manifold $\MM$ one should also require that the anomaly functional is suitably invariant under coupling-constant redefinitions. This can be achieved by utilising a connection $\nabla$ on the bundle of operators. For exactly marginal couplings, the WZ consistency condition on the $\MM$-covariantised version of the anomaly implies that the connection $\nabla$ is compatible with the Zamolodchikov metric, \cite{Gomis:2015yaa}. In this paper, we examine whether this argument can be extended beyond the case of the exactly marginal operators.  

Since the presence of a contact term like the one in \eqref{extractingHB}, in any phase of the theory, has been established independently by the analysis of Ward identities, in all phases the anomaly functional includes a term of the form
\begin{align}
  \label{eq:29}
  \delta_\sigma W \propto \int d^4x \sqrt \gamma \delta\sigma [ G_{I\bar J} \lambda^I \Box^n \bar{\lambda}^J + \ldots ]\;,
\end{align}
where $G_{IJ}$ are the corresponding anomaly coefficients. This term should be covariantised on the corresponding vector bundle of operators over $\MM$. We perform this covariantisation for operators of scaling dimension $\Delta = 3,4,5$ in Sec.~\ref{main1} and show that the WZ consistency condition \eqref{eq:11} requires that the anomaly is covariantly constant. In the case of scaling-dimension 4 operators the arguments of Refs\ \cite{Jack:1990eb,Jack:2013sha,Gomis:2015yaa} are modified to capture the properties of marginal, but not necessarily exactly marginal operators. Our analysis is completely general and does not employ supersymmetry at any stage. We expect that similar arguments can be applied to all higher values of integer scaling dimension $\Delta$, but the anomaly functional becomes significantly more complicated with increasing $\Delta$. Indeed, already at $\Delta=5$ we present WZ-consistent anomaly functionals, which contain hundreds of terms in the flat-space limit. We notice that in both the cases of $\Delta=4, 5$ anomalies, new terms in the anomaly functional that involve the curvature of the corresponding operator bundles are crucial in order to satisfy the WZ consistency conditions.

The case of $\Delta=2$ operators is special and requires a separate discussion: the anomaly functional is automatically WZ-consistent and \eqref{eq:11} does not lead to further restrictions. To make a non-trivial statement, we need to use the $\NN=2$ supersymmetry to relate the $\Delta=2$ anomaly to the anomaly of the exactly marginal operators. An argument in favour of this relation is sketched in Sec.\ \ref{delta2} alongside an explicit tree-level check for $\NN=2$ SCQCD in the conformal and Higgs phases.

\section{WZ Consistency Conditions in 4D CFTs}\label{main1}

We follow closely the discussion and notation of references \cite{Schwimmer:2018hdl,Schwimmer:2019efk}. $W=\log Z$ is the generating functional of correlation functions. It is a functional of the spacetime-dependent sources (couplings). In this section, we focus on four spacetime dimensions and type-B conformal anomalies of scalar operators. Such anomalies exist when the operators have scaling dimensions $\Delta = 2+n$ with $n\in \mathbb N_0$.

We will denote the operators of interest as $\OO_I$ and their corresponding sources as $\lambda^I$. Note that although we ultimately have $\mathcal N=2$  applications in mind, we will use a real basis of operators and will not require supersymmetry for any of the arguments presented in this section. When the operators are exactly marginal they will be denoted as $\Phi_i$ and their corresponding couplings as $\lambda^i$. Clearly, the index $i$ takes values up to the dimension of the conformal manifold $\mathcal M$. The more general indices $I$ label conformal primary operators in a sub-bundle of operators of fixed integer dimension and the corresponding conformal anomalies will be denoted $G_{I J}$. The background spacetime metric will be denoted $\gamma_{\mu\nu}$ with Greek letters reserved for the spacetime coordinate indices. Vector bundles over the conformal manifold can be equipped with a connection. For a discussion of this connection in the context of conformal perturbation theory see \cite{Ranganathan:1993vj,Papadodimas:2009eu}. For a related discussion in radial quantization see \cite{Baggio:2017aww}. The components of the connection on the sub-bundle of $\OO_I$ operators will be denoted as $(A_i)^I_J$, whereas the connection on the tangent space of $\Phi_i$ operators as $\Gamma_{ij}^k$. The corresponding covariant derivative on the conformal manifold will be denoted as $\nabla_i$.

In this section we follow the general strategy of \cite{Jack:1990eb,Jack:2013sha,Gomis:2015yaa}, where the basic ansatz for the Weyl variation of $W$ was covariantised not only in spacetime but also in the tangent bundle of the conformal manifold $T\mathcal M $. Accordingly, for the type-B Weyl anomalies of exactly marginal operators \cite{Jack:1990eb,Jack:2013sha,Gomis:2015yaa} proposed the anomaly functional:\footnote{For the case of a single coupling, this expression is related to the Fradkin--Tseytlin--Paneitz--Riegert operator \cite{Fradkin:1981jc, Fradkin:1982xc,Riegert:1984kt,Paneitz_2008}. A six-dimensional generalisation of this operator was presented in \cite{Osborn:2015rna}.}
\begin{align}
  \label{eq:30}
  \delta_\sigma W \propto &\int d^4 x \sqrt{\gamma} \delta\sigma \,  G_{ij}\bigg( (\square \lambda^{i}+\Gamma_{kl}^{i} \partial^{\mu} \lambda^{k} \partial_{\mu} \lambda^{l})(\square \lambda^{j}+\Gamma_{mn}^{j} \partial^{\nu} \lambda^{m} \partial_{\nu} \lambda^{n})\cr
  &\qquad \qquad \qquad \qquad \qquad \qquad -2 \pd_\mu \lambda^i (R^{\mu\nu} - \frac{1}{3} \gamma^{\mu\nu} R) \pd_\nu \lambda^j \bigg)
~,
\end{align}
with $R_{\mu\nu}$ and $R$ the spacetime Ricci tensor and scalar respectively. Clearly, this functional is sensitive only to the symmetric part $\Gamma_{(kl)}^{i}$ of the connection. The WZ-condition identifies it to be the Levi--Civita connection on $\mathcal M$, and under the further assumption that the connection is torsion-free \cite{Ranganathan:1993vj} 
one obtains that the anomaly is covariantly constant on $\mathcal M$,
\begin{align}\label{eq:marginal}
    \nabla_i G_{jk}= 0\,.
\end{align}

This approach can be generalised to generic operators $\OO_I$, where covariantisation on the conformal manifold translates into the invariance of $\delta_\sigma W$ under a change of basis in the vector space of $\OO_I$s. We discover that imposing the WZ consistency condition will typically lead to $\nabla_i G_{IJ} = 0$.

We emphasise that this result is independent of the phase of the theory. The anomaly functional can be understood as the local Weyl variation of the generating functional $W$ with appropriate boundary conditions for the fields. The infinitesimal local Weyl parameters, $\delta\sigma(x)$, by definition vanish at the boundary of spacetime and parametrise transformations that are valid both in the conformally symmetric and broken phases \cite{Schwimmer:2010za}. Moreover, the asymptotic behaviour of $\delta\sigma(x)$ also guarantees that any boundary terms that involve $\delta\sigma(x)$ (obtained after integration by parts) can be safely ignored in the upcoming discussion. 

We will now summarise the key ingredients of the calculation, before specialising to type-B anomalies for operators with $\Delta = 3,4,5$.\footnote{Here the $\Delta = 4$ case refers exclusively to Weyl anomalies for marginal operators that are not exactly marginal---they can be marginally relevant or irrelevant.} The $\Delta = 2$ case cannot be constrained with a simple analysis of the Wess-Zumino consistency condition and will be treated separately in Sec.~\ref{delta2}. The expressions  $\delta_\sigma W$ for cases with a single source can be found in \cite{Schwimmer:2019efk} and form the starting point of our discussion. We study the WZ consistency conditions after we covariantise the expressions in Ref.\ \cite{Schwimmer:2019efk} with respect to the conformal manifold. In the process we discover that a fully covariant anomaly functional requires new terms that have not appeared previously in the literature.

\subsection{Covariantisation on the Conformal Manifold}

In what follows we will make an important distinction between the exactly marginal couplings $\lambda^i$ that parametrise the conformal manifold and the remaining non-exactly marginal sources $\lambda^I$. Geometrically, the couplings $\lambda^i$ are, in general, non-linear coordinates on the curved conformal manifold, which are allowed to also depend non-trivially on the spacetime coordinates. Equivalently, we view the spacetime derivatives $\partial_\mu \lambda^i$ as components on the tensor product of the spacetime cotagent bundle and the conformal manifold tangent bundle. The couplings $\lambda^I$ are viewed, instead, as sections of a vector bundle. They can depend both on the spacetime and conformal manifold coordinates.

Accordingly, under a change of basis on the tangent bundle of the conformal manifold
\begin{align}
    \label{margTransf}
    \partial_\mu \lambda^i = \frac{\partial \lambda^i}{\partial {\lambda'}^{j'}} \partial_\mu {\lambda'}^{j'}\;.
\end{align}
On the other hand, under a change of basis on each fibre of the $\lambda^I$-vector bundle
\begin{align}
  \label{eq:1}
  \lambda^I = \frac{\pd \lambda^I}{\pd {\lambda'}^{I'}} {\lambda'}^{I'}\;,
\end{align}
where the transformation matrix $\frac{\pd \lambda^I}{\pd {\lambda'}^{I'}}$ depends on the $\lambda^i(x^\mu)$ only. As a result, we define covariant derivatives on the conformal manifold in terms of the connection components $(A_i)^I_J$ as
\begin{align}
  \label{eq:6}
 \nabla_i \lambda^I = \pd_i \lambda^I + (A_i)^I_J \lambda^J\;.
\end{align}
The generalised covariant derivative is then naturally given by\footnote{Here we are explicitly stressing that $\nabla_\mu$ and $\partial_\mu$ have to be understood at fixed $\lambda^i$, but later this will be left implicit.}
\begin{align}
  \label{eq:7}
 \widehat \nabla_\mu \lambda^I:= \nabla_\mu \lambda^i  \nabla_i\lambda^I + \nabla_\mu \lambda^I |_{\lambda^i = {\rm fixed}}=\pd_\mu \lambda^i  \nabla_i\lambda^I + \partial_\mu \lambda^I |_{\lambda^i = {\rm fixed}}\;,
\end{align}
with $\nabla_\mu $ the standard spacetime-covariant derivative.

Compared to the unhatted differential operators used in \cite{Schwimmer:2019efk}, commutators of our hatted operators can lead to curvature terms on $\mathcal M$. The latter can be easily evaluated by using the definition of the generalised covariant derivative and the fact that $\partial_\mu (A_i)^I_J |_{\lambda^i = {\rm fixed}}=0$, i.e. 
\begin{align}
  \label{eq:34}
  (F_{\mu\nu})^I_J \lambda^J :=& [\widehat\nabla_\mu, \widehat\nabla_\nu]\lambda^I = \partial_{\mu} \lambda^{i} \partial_{\nu} \lambda^{j} (F_{ij})^I_J \lambda^J\;,
\end{align}
where $(F_{ij})^I_J=\partial_{i}\left(A_{j}\right)_{J}^{I}-\partial_{j}\left(A_{i}\right)_{J}^{I}+\left(A_{i}\right)_{K}^{I}\left(A_{j}\right)_{J}^{K}-\left(A_{j}\right)_{K}^{I}\left(A_{i}\right)_{J}^{K} $.

Under the change of basis \eqref{margTransf}-\eqref{eq:1}, the connection transforms inhomogeneously as
\begin{align}
  \label{eq:3}
  (A_i)^I_J = \frac{\pd {\lambda'}^{i'}}{\pd \lambda^i}\frac{\pd {\lambda'}^{J'}}{\pd \lambda^J}\frac{\pd \lambda^I}{\pd {\lambda'}^{I'}}(A_{i'})^{I'}_{J'} - \frac{\pd {\lambda'}^{i'}}{\pd \lambda^i}\frac{\pd {\lambda'}^{J'}}{\pd \lambda^J}\frac{\pd^2 \lambda^I}{\pd {\lambda'}^{i'}\pd {\lambda'}^{J'}}
\end{align}
such that
\begin{align}
  \label{eq:31}
\nabla_{i} {\lambda}^{I} =\frac{\partial {\lambda'}^{i'}}{\partial {\lambda}^{i}} \frac{\pd  {\lambda}^{I}}{\pd {\lambda'}^{I'}}\nabla_{i'} {\lambda}'^{I'}\;,
\end{align}
which in turn implies
\begin{align}
  \label{eq:4}
   \widehat \nabla_\mu \lambda^I  =  \frac{\pd {\lambda}^{I}}{\pd {\lambda'}^{I'}}\widehat \nabla_\mu{ \lambda'}^{I'} \;.
\end{align}
Therefore, standard differential operators can be covariantised on $\mathcal{M}$ by upgrading the usual spacetime-covariant derivative $\nabla_\mu$ to $ \widehat \nabla_\mu$. For example, the $\mathcal{M}-$covariant Laplacian $\widehat \Box $, which reads 
\begin{align}
  \label{eq:32}
  \widehat \Box \lambda^I&:= \widehat \nabla_\mu \widehat \nabla^\mu \lambda^I\cr
                           &=\partial^\mu\lambda^i\nabla_i\left(\widehat{\nabla}_\mu \lambda^I\right)+\nabla^\mu\Big|_{\lambda^i \text{fixed}}\widehat{\nabla}_\mu \lambda^I\\
  &=\partial^\mu\lambda^i\left(\partial_i\widehat{\nabla}_\mu \lambda^I+(A_i)^I_J\widehat{\nabla}_\mu \lambda^J\right)+\partial^\mu\Big|_{\lambda^i \text{fixed}} \widehat{\nabla}_\mu \lambda^I-g^{\mu\nu}\Gamma^\rho_{\nu\mu}\widehat{\nabla}_\rho \lambda^I\;,\nonumber
\end{align}
transforms as
\begin{align}
  \label{eq:5}
  \widehat \Box \lambda^I = \frac{\pd \lambda^I}{\pd {\lambda'}^{I'}}\widehat \Box \lambda'^{I'}\;.
\end{align}
As a result, to get anomaly functionals invariant under a change of basis in the space of $\OO$s, one can consider the ones written in \cite{Schwimmer:2019efk} and simply replace all spacetime covariant derivatives with their hatted versions. However, because of \eqref{eq:34} this minimal prescription is, in general, sensitive to ordering choices and does not guarantee WZ consistency.

We conclude this section with some remarks on $\lambda^i$ and by explicitly stressing how our framework is compatible with the one of \cite{Jack:1990eb,Jack:2013sha,Gomis:2015yaa}. As the exactly marginal coupling $\lambda^i$ is not a tensor (it is a coordinate on the conformal manifold), the generating functional cannot display an explicit $\lambda^i$ dependence. Instead, the anomaly can depend on it only through its infinitesimal variation, i.e.
\begin{equation}
\lambda^i_\mu:=\partial_\mu\lambda^i=\nabla_\mu\lambda^i\,.
\end{equation}
This object serves as a pullback from the conformal manifold to spacetime, which could have been appreciated already at the level of formula \eqref{eq:7}. It has good transformation properties \eqref{margTransf} and can then be acted upon by the generalised covariant  derivative:
\begin{align}
    \widehat \nabla_\mu \lambda^{i}_{\nu}&=
    \nabla_\mu \lambda^{i}_\nu+\lambda^{j}_\mu\Gamma^i_{jk}\lambda^{k}_\nu\;.
\end{align}
Thus, within our framework, \eqref{eq:30} can be more succinctly recast into the form
\begin{align}
  \label{eq:30bis}
  \delta_\sigma W \propto &\int d^4 x \sqrt{\gamma} \delta\sigma \,  G_{ij}\bigg( \widehat \nabla_\mu \lambda^{i\mu}\widehat \nabla_\nu \lambda^{j\nu}  -2  \lambda^i_\mu (R^{\mu\nu} - \frac{1}{3} \gamma^{\mu\nu} R) \lambda^j_{\nu} \bigg)\cr
  =&\int d^4 x \sqrt{\gamma} \delta\sigma \, \bigg( \widehat \nabla_\mu \lambda^{i\mu}\widehat \nabla_\nu \lambda_i^{\nu}  -2  \lambda^i_\mu (R^{\mu\nu} - \frac{1}{3} \gamma^{\mu\nu} R) \lambda_{i\nu} \bigg)
~.
\end{align}
In the second line we have implicitly used the fact that $\Gamma^i_{jk}$ is given by the Christoffel symbol (so that $\widehat \nabla_\mu G_{jk}=0$). The fact that $\Gamma^i_{jk}$ is symmetric yields many simplifications, e.g.
\begin{align}\label{simplification}
    \widehat \nabla_{[\mu} \lambda^{i}_{\nu]}=0\;,\qquad
    \lambda^{i}_{[\mu}\nabla_i \lambda^{j}_{\nu]}=0\;,
\end{align}
where the first equation guarantees that the Bianchi identity $\nabla_{[i}(F_{jk]})^I_J=0 $ gets pulled-back onto $\widehat \nabla_{[\mu}(F_{\nu\rho]})^I_J=0 $.

\subsection{Weyl Transformations}

In four spacetime dimensions, an infinitesimal local Weyl transformation acts on the spacetime metric $\gamma_{\mu\nu}$ as
\begin{align}
  \label{eq:22}
  \delta_\sigma \gamma_{\mu\nu} &= 2 \delta\sigma\;\gamma_{\mu\nu}\,.
\end{align}
The Christoffel symbols, the Ricci tensor $R_{\mu\nu}$ and the Ricci scalar $R$ transform accordingly
\begin{align}
  \label{eq:23}
       \delta_\sigma  \Gamma_{\mu\nu}^{\rho}&=\gamma^{\rho\sigma}\left(\gamma_{\nu\sigma}\partial_\mu\delta\sigma+\gamma_{\mu\sigma}\partial_\nu\delta\sigma-\gamma_{\mu\nu}\partial_\sigma\delta\sigma\right)\cr
 \delta_\sigma R_{\mu\nu} &= -2 \nabla_\mu \nabla_\nu \delta\sigma -\gamma_{\mu\nu}\Box \delta\sigma\\
\delta_\sigma R &= -2 \delta\sigma\;R - 6 \Box \delta\sigma\;. \nonumber          
\end{align}
For an operator of conformal scaling dimension $\Delta$, one has classically $ \delta_\sigma \OO_I= - \Delta \OO_I \delta \sigma$. Thus,
\begin{align}
  \label{eq:24}
  \delta_\sigma \lambda^I  = (\Delta - 4) \delta \sigma\; \lambda^I  ~, ~~
  \delta_\sigma \lambda^i = 0\;.  
\end{align}
Being a number, the anomaly has vanishing classical dimension, so
\begin{align}
  \label{eq:8}
     \delta_\sigma G_{IJ}  = 0\;,
\end{align}
while the uniform Weyl variation of $\nabla_i \lambda^I$ leads to
\begin{align}
  \label{eq:25}
                           \delta_\sigma (A_i)^I_J &= 0\;.                                                    
\end{align}
One then finds that standard equations such as
\begin{align}
  \label{eq:9}
                                 \delta_\sigma \pd_\mu \lambda^I & = (\Delta - 4) \pd_\mu \lambda^I \delta\sigma + (\Delta-4) \pd_\mu \delta\sigma\; \lambda^I\\
                                   \delta_\sigma \Box \lambda^I & = (\Delta - 6) \delta\sigma\; \Box \lambda^I + 2(\Delta -3) \pd_\mu \delta\sigma \;\pd^\mu \lambda^I + (\Delta - 4) \lambda^I \Box \delta\sigma\;.
\end{align}
can be straightforwardly extended to
\begin{align}
  \label{eq:10}
                                 \delta_\sigma \widehat\nabla_\mu \lambda^I & = (\Delta - 4) \widehat\nabla_\mu \lambda^I \delta\sigma + (\Delta-4) \pd_\mu \delta\sigma\; \lambda^I\\
                                   \delta_\sigma \widehat\Box \lambda^I & = (\Delta - 6) \delta\sigma\; \widehat\Box \lambda^I + 2(\Delta -3) \pd_\mu \delta\sigma \;\widehat\nabla^\mu \lambda^I + (\Delta - 4) \lambda^I \Box \delta\sigma\;.
\end{align}
and accordingly for quantities with raised spacetime indices. These expressions will 
be useful in the calculations that we will be performing below. 

\subsection{$\Delta=3$ Operators}

We begin the construction of fully covariant and WZ-consistent anomaly functionals with the case of $\Delta=3$. According to the discussion around Eq.~\eqref{eq:29}, the ansatz for this case should contain two derivatives. In order to address the Weyl-cohomological problem, we will first characterise terms in the anomaly functional that are cohomologically trivial. We start with the following expression for the generating functional of connected correlation functions
\begin{align}
    W^{\rm exact}&=\int d^4 x \sqrt{\gamma}\, \Bigg[ G_{IJ}  \widehat\nabla^\mu \lambda^I \widehat \nabla_\mu \lambda^J  + A_1\lambda^I \widehat \Box \lambda^JG_{IJ}+A_2 G_{IJ}   \lambda^I \lambda^J R\cr
 &\qquad \qquad \qquad \qquad+A_3\lambda^{I} \widehat{\nabla}^{\mu} \lambda^{J} \widehat{\nabla}_{\mu} G_{I J}+A_4 \lambda^{I} \lambda^{J}  \widehat{\Box} G_{I J}
    \Bigg]\;.
\end{align}
By computing its Weyl variation, and after integrating by parts, we find that the most general exact (i.e. cohomologically trivial) anomaly functional is
\begin{align}
\label{cohotrivial}
    \delta_\sigma W^{\rm  exact}\propto\int d^4 x \sqrt{\gamma} \delta\sigma\Bigg[&2(-1+A_1+6A_2) G_{IJ}  \widehat\nabla^\mu \lambda^I \widehat \nabla_\mu \lambda^J +2(-1+A_1+6A_2)\lambda^I \widehat \Box \lambda^JG_{IJ}\cr
    & \qquad +2(-1+2A_1+12A_2-A_3+2A_4)\lambda^{I} \widehat{\nabla}^{\mu} \lambda^{J} \widehat{\nabla}_{\mu} G_{I J}\cr
    & \qquad +(A_1+6A_2-A_3+2A_4) \lambda^{I} \lambda^{J}  \widehat{\Box} G_{I J}\Bigg]\;.
\end{align}
From the above one can deduce that:
\begin{itemize}
    \item[$(a)$] An anomalous Weyl generating functional containing $\delta \sigma G_{IJ}   \lambda^I \lambda^J R $ cannot be cohomologically trivial.
    \item[$(b)$] The term $\lambda^{I} \lambda^{J}  \widehat{\Box} G_{I J}$ is cohomologically equivalent to $\lambda^{I} \widehat{\nabla}^{\mu} \lambda^{J} \widehat{\nabla}_{\mu} G_{I J}$.
    \item[$(c)$] The term $\lambda^{I} \widehat{\nabla}^{\mu} \lambda^{J} \widehat{\nabla}_{\mu} G_{I J}$ is cohomologically equivalent to $G_{IJ}  \widehat\nabla^\mu \lambda^I \widehat \nabla_\mu \lambda^J  + \lambda^I \widehat \Box \lambda^JG_{IJ}$ and by going to momentum space, one sees that the latter does not contribute to the anomaly.
\end{itemize}
Hence, modulo cohomologically trivial terms and up to integration by parts, the most general Weyl anomalous functional is given by
\begin{align}
    \delta_\sigma W=\int d^4 x \sqrt{\gamma}\, \delta \sigma  G_{IJ}\Bigg[  C_1 \widehat\nabla^\mu \lambda^I \widehat \nabla_\mu \lambda^J  + C_2\lambda^I \widehat \Box \lambda^J+C_3  \lambda^I \lambda^J R \Bigg]
    \label{first}
\end{align}
with $C_1\not=C_2$. Imposing the WZ consistency condition leads to the following independent solutions for the anomaly functional:
\begin{align}\label{w1}
     \delta_\sigma W^{(1)}&=\int d^4 x \sqrt{\gamma}\, \delta \sigma  G_{IJ}\Bigg[\widehat\nabla^\mu \lambda^I \widehat \nabla_\mu \lambda^J +\frac16  \lambda^I \lambda^J R \Bigg]\,,\\
\label{w2}    \delta_\sigma W^{(2)}&=\int d^4 x \sqrt{\gamma}\, \delta \sigma  G_{IJ}\Bigg[ \lambda^I \widehat \Box \lambda^J-\frac16  \lambda^I \lambda^J R \Bigg]\;.
\end{align}
For $ \delta_\sigma W^{(1)}$ one needs to impose $\nabla_i G_{IJ}=0$, while $ \delta_\sigma W^{(2)}$ is automatically WZ consistent.\footnote{The WZ consistency condition imposes $C_3= \frac{1}{6}(C_1 - C_2)$ so the most general anomaly functional is given by $\delta_\sigma W=C_1 \delta_\sigma W^{(1)}+C_2 \delta_\sigma W^{(2)}$ with $C_1\not=C_2$. Terms with $C_1=C_2$ cannot capture the anomaly, see point $(c)$ above.} It is interesting to observe that \eqref{w1} and \eqref{w2} are equivalent upon integration by parts when $\nabla_i G_{I J}=0$, leading to a self-consistent picture.

\subsection{$\Delta=4$ Operators}

The classical Weyl variations \eqref{eq:10} do not distinguish between the exactly-marginal couplings $\lambda^i$ and the marginally relevant or irrelevant $\lambda^I$. However, in our formalism these two sets of couplings are treated differently---the $\lambda^i$ are non-linear coordinates on the conformal manifold but the $\lambda^I$ are linear coordinates on a vector bundle. Accordingly, in the conformal phase, we can interpret the anomalies $G_{ij}$ as a Zamolodchikov metric on the conformal manifold, but the anomalies $G_{IJ}$ do not have such an interpretation. This will soon translate to a different type of anomaly functional for the anomalies $G_{IJ}$, which is sensitive to the curvature of the corresponding operator bundles. Examples of theories with non-exactly marginal $\Delta=4$ operators, whose curvature is non-trivial, are abundant in 4D $\NN=2$ SCFTs, see e.g.\ \cite{Baggio:2014ioa}.  

It is sensible to start with an anomaly functional, which is similar to \eqref{eq:30bis} for the exactly marginal operators 
\beq
\label{wz4ca}
\delta_\sigma W \propto \int d^4 x \sqrt{\gamma} \delta\sigma \,  G_{IJ}\bigg[ \widehat \Box \lambda^I \widehat \Box \lambda^J -2 \widehat\nabla_\mu \lambda^I (R^{\mu\nu} - \frac{1}{3} \gamma^{\mu\nu} R) \widehat\nabla_\nu \lambda^J \bigg]
~.
\eeq
For exactly marginal operators $\widehat\nabla_{[\mu}\lambda^i_{\nu]}=0$ from \eqref{simplification}. Instead, for non-exactly marginal $\Delta = 4$ operators  $[\widehat\nabla_{\mu},\widehat\nabla_{\nu}]\lambda^I=(F_{\mu\nu})^I_J\lambda^J$. As a result, we expect that terms containing either $(F_{\mu\nu})^K_J$ or explicit $(F_{ij})^K_J$ contributions will mark a distinctive difference compared to the exactly-marginal case. Indeed, when checking the WZ-consistency condition for \eqref{wz4ca}, one finds that
\begin{align}
 \label{eq:33}
 \delta_{\sigma_{[2}}\delta_{\sigma_{1]}}W\propto &\int d^4x\sqrt{\gamma}\delta \sigma_{[1}\partial^\nu\delta\sigma_{2]}\times\cr
  &\qquad\times\widehat{\nabla}^\mu \lambda^I\left[-4\widehat\nabla_\mu G_{IJ}\widehat\nabla_\nu\lambda^J+2\widehat\nabla_\nu G_{IJ} \widehat\nabla_\mu\lambda^J-4G_{IK}\lambda^J (F_{\mu\nu})^K_J\right] ~.  
\end{align}
The expression on the RHS does not vanish automatically even after imposing $\nabla_i G_{IJ}=0$: extra terms need to be added to \eqref{wz4ca} to cancel the last term in \eqref{eq:33}. One can exhaustively prove that terms constructed out of $(F_{\mu\nu})^I_J$ are closed with respect to the Weyl-cohomology and cannot achieve the desired goal. We are thus forced to use terms where $(F_{ij})^I_J$ appears explicitly and does not combine with pullbacks to give $(F_{\mu\nu})^I_J=(F_{ij})^I_J\lambda^i_\mu\lambda^j_\nu$. We notice, using the first equation in \eqref{eq:23}, that $\delta_\sigma(\widehat\nabla_{\rho}\lambda^i_\nu)\sim \partial_{\rho}\delta \sigma \lambda^i_\nu$, and as a result
\begin{equation}
    \delta_\sigma\left((F_{ij})^J_K\lambda^j_\rho \widehat{\nabla}_\mu \lambda^i_\nu\right)=(F_{ij})^J_K\lambda^j_\rho \delta_\sigma\left( \widehat{\nabla}_\mu \lambda^i_\nu \right)\sim (F_{\rho\nu})^J_K \partial_\mu \delta \sigma \quad.
\end{equation}
We are thus led to consider terms with the schematic structure: $G_{IJ}(F_{ij})^I_K\lambda^j_\rho\lambda^K \widehat{\nabla}_\mu\lambda^i_\nu\widehat{\nabla}_\sigma\lambda^J$. By taking into account all possible contractions for the spacetime indices, we arrive at the generating functional
\begin{align}
\label{functionalJes}
    \delta_\sigma W \propto \int d^4 x & \sqrt{\gamma} \delta\sigma \,  G_{IJ}\bigg[ \widehat \Box \lambda^I \widehat \Box \lambda^J -2 \widehat\nabla_\mu \lambda^I (R^{\mu\nu} - \frac{1}{3} \gamma^{\mu\nu} R) \widehat\nabla_\nu \lambda^J+ \nonumber\\
    & +(F_{ij})^I_K(E_3g^{\mu \sigma}g^{\nu \rho}+E_2 g^{\mu \rho}g^{\nu \sigma}+E_1g^{\mu \nu}g^{\rho \sigma})\lambda^j_\rho \widehat{\nabla}_\mu\lambda^i_\nu\lambda^K\widehat{\nabla}_\sigma\lambda^J \bigg],
\end{align}
where $E_1,E_2,E_3$ are free constants. The WZ consistency condition can be satisfied by setting $\nabla_i G_{IJ}=0$ and $E_1+E_2+E_3=-2$. The fact that only the combination $E_1+E_2+E_3=-2$ survives the WZ condition suggests a relation between the three terms in the second line of \eqref{functionalJes}. Indeed, the terms parametrised by $E_2$ and $E_3$ are identical as a consequence of the identity $\widehat\nabla_{[\mu}\lambda^i_{\nu]}=0$. This leaves a single combination in \eqref{functionalJes}---the difference between the terms parametrised by $E_1$ and $E_2$ being closed, but not exact. The resultant anomaly functional \eqref{functionalJes} is the WZ-consistent functional that captures the type-B anomalies $G_{IJ}$ for non-exactly-marginal $\Delta=4$ operators.

We can draw two lessons from this discussion. First, we verify once again that the condition $\nabla_i G_{IJ}=0$ is necessary to obtain WZ consistency. Second, and on a more technical level, we notice that in order to cancel $F_{\mu\nu}$-terms in the WZ consistency condition \eqref{eq:33}, one needs to add to the generating functional terms where $F_{ij}$ factors come contracted with (differentiated) pull-backs. The specific terms added in \eqref{functionalJes} contributed to the WZ condition only with $F_{\mu\nu}$ combinations. It turns out that this is a special feature of $\Delta=4$ operators (for which both $\delta_\sigma\lambda^I$ and $\delta_\sigma \widehat{\nabla}\lambda^I$ vanish). In the next section, we will see that $F_{ij}$-terms provide contributions to the WZ condition of $\Delta=5$ anomalies that do not combine to produce $F_{\mu\nu}$. This feature will add to the complexity of the $\Delta=5$ anomaly functionals. 

\subsection{$\Delta=5$ Operators}

The Osborn equation for type-B anomalies of irrelevant operators in even spacetime dimensions is subtle. Its intricacies were discussed in \cite{Schwimmer:2019efk}, the main lesson being that in order to ensure the consistency of the anomalous part, one has to introduce a beta function for the spacetime metric. We will generalise the analysis of  \cite{Schwimmer:2019efk} to the case of multiple irrelevant sources $\lambda^I$, starting with the most general ansatz for the spacetime Weyl variation $\delta_\sigma \gamma_{\mu\nu}$ that is quadratic in the sources $\lambda^I$. One needs to first impose that $\delta_{[\sigma_2}\delta_{\sigma_1]}\gamma_{\mu\nu}=0$ and then remove the cohomologically trivial terms from $\delta_\sigma\gamma_{\mu\nu}$.\footnote{The latter are those solutions $(\delta_\sigma\gamma_{\mu\nu})_{\rm trivial}$ to $\delta_{[\sigma_2}\delta_{\sigma_1]}\gamma_{\mu\nu}=0$ that can be written as $(\delta_\sigma\gamma_{\mu\nu})_{\rm trivial}=\delta_\sigma \widehat\gamma_{\mu\nu}-2\delta\sigma \widehat\gamma_{\mu\nu}$ for a metric $\widehat\gamma_{\mu\nu}$. Therefore the redefined metric $\gamma_{\mu\nu}\mapsto\gamma_{\mu\nu}-\widehat\gamma_{\mu\nu}$, continues to transform classically.} The outcome of this analysis, at quadratic order in the sources, is that the variation of the metric $\delta_\sigma \gamma_{\mu\nu}$ is essentially the covariantised version of the one proposed by \cite{Schwimmer:2019efk}, i.e.
\begin{align}
  \label{eq:12}
  \delta_{\sigma} \gamma_{\mu \nu}=2 \delta\sigma \;\gamma_{\mu \nu}+&\alpha \delta\sigma G_{I J}\left(R_{\mu \nu} \lambda^I\lambda^J+2\lambda^I  \widehat \nabla_{(\mu} \widehat \nabla_{\nu)} \lambda^J -3 \gamma_{\mu \nu} \widehat \nabla^\rho\lambda^I\widehat \nabla_\rho\lambda^J +\gamma_{\mu \nu}\lambda^I \widehat\square \lambda^J\right)\cr
  +&\beta \delta\sigma\gamma_{\mu\nu}\left(R \lambda^I\lambda^J+6\lambda^I\widehat\square \lambda^J-12\widehat \nabla^\rho\lambda^I\widehat \nabla_\rho\lambda^J \right)+ O(\lambda^4)\;,
\end{align}
where $\alpha$ and $\beta$ are free parameters. Here we have neglected---already at $O(\lambda^2)$---terms that vanish when $\nabla_i G_{I J}=0$; one can prove that they sit in a cohomology class different to that of the ones proportional to $G_{IJ}$, hence their presence would not modify \eqref{eq:12}. Moreover, such terms will not play a role in the computations that we will display below.

As a starting point for the analysis of the $\Delta = 5$ anomaly functional, we consider the covariantised version of the expression derived in \cite{Schwimmer:2019efk}, which to quadratic order in the irrelevant sources reads\footnote{We thank M.~Broccoli for pointing out a missing factor of $\frac{1}{2}$ between the $C_{\mu \nu \rho\sigma} C^{\mu \nu \rho \sigma}$ and $\lambda^2$ terms in \cite{Schwimmer:2019efk}. This factor can also be confirmed by an independent holographic computation \cite{Broccoli:2021icm}.}
\begin{align}
  \label{eq:14}
\mathcal{A}&= c\; C_{\mu \nu \rho\sigma} C^{\mu \nu \rho \sigma}+\frac{c}{2}\alpha G_{IJ}\left\{\widehat\square \lambda^I \widehat\square^{2} \lambda^J-\frac{13}{8} R R^{\mu \nu} R_{\mu \nu} \lambda^I\lambda^J+\frac{53}{162} R^{3} \lambda^I\lambda^J+\frac{4}{3} R^{\mu \nu} R^{\lambda\sigma} R_{\mu \lambda \nu \sigma} \lambda^I\lambda^J\right.\cr
&-\frac{1}{8} R R_{\mu \nu \lambda \sigma} R^{\mu \nu \lambda \sigma} \lambda^I\lambda^J+\frac{43}{72} R_{\mu \nu \lambda \sigma} R^{\mu \nu \alpha \beta} R_{\alpha \beta}{ }^{\lambda \sigma} \lambda^I\lambda^J-\frac{35}{72} R^{2} \lambda^I \widehat\square \lambda^J+\frac{25}{24} R_{\mu \nu \lambda \sigma} R^{\mu \nu \lambda \sigma} \lambda^I \widehat\square \lambda^J \cr
            &-\frac{1}{36} \nabla^{\mu} R \nabla_{\mu} R \lambda^I\lambda^J+\frac{167}{12} R^{\mu \nu} R_{\mu \nu} \widehat\nabla^{\alpha} \lambda^I \widehat\nabla_{\alpha} \lambda^J-\frac{101}{24} R^{2} \widehat\nabla^{\alpha} \lambda^I \widehat\nabla_{\alpha} \lambda^J\cr
&-\frac{79}{24} R^{\mu \nu \lambda \sigma} R_{\mu \nu \lambda \sigma} \widehat\nabla^{\alpha} \lambda^I\widehat \nabla_{\alpha} \lambda^J-\frac{1}{3} R \widehat\square \widehat\nabla^{\mu} \lambda^I \widehat\nabla_{\mu} \lambda^J-\frac{10}{9} R^{\mu \nu} \nabla_{\mu} \nabla_{\nu} R \lambda^I\lambda^J+\frac{7}{9} R^{\mu \nu} R \lambda^I \widehat\nabla_{\mu} \widehat\nabla_{\nu} \lambda^J\cr
&+\frac{1}{36} \square R\lambda^I \widehat\square \lambda^J-\frac{16}{9} R\widehat\square \lambda^I\widehat\square \lambda^J+\nabla^{\mu} R \widehat\nabla_{\mu} \lambda^I \widehat\square \lambda^J+\frac{1}{6} R \lambda^I \widehat\square^{2} \lambda^J-4 R^{\mu \nu} \widehat\nabla_{\mu} \lambda^I \widehat\square \nabla_{\nu} \lambda^J\cr
&-\frac{37}{18} R_{\mu \nu} \nabla^{\mu} R \lambda^I \widehat\nabla_{\nu} \lambda^J-22 R_{\mu}^{\alpha} R_{\nu \alpha} \widehat\nabla^{\mu} \lambda^I \widehat\nabla^{\nu} \lambda^J+\frac{116}{9} R^{\mu \nu} R \widehat\nabla^{\mu} \lambda^I \widehat\nabla^{\nu} \lambda^J\cr
&-13 R^{\alpha \beta} R_{\mu \alpha \nu \beta} \widehat\nabla^{\mu} \lambda^I \widehat\nabla^{\nu} \lambda^J-\frac{5}{18} \nabla^{\mu} \nabla^{\nu} R \lambda^I \widehat\nabla_{\mu} \widehat\nabla_{\nu} \lambda^J-\frac{5}{9} R \widehat\nabla_{\mu}\widehat \nabla_{\nu} \lambda^I \widehat\nabla^{\mu} \widehat\nabla^{\nu} \lambda^J\cr
&-5 R^{\beta \gamma} \nabla_{\gamma} R_{\alpha \beta} \lambda^I \widehat\nabla^{\alpha} \lambda^J-\frac{8}{3} R_{\alpha}^{\gamma} R^{\alpha \beta} \lambda^I \widehat\nabla_{\beta} \widehat\nabla_{\gamma} \lambda^J+\frac{10}{3} R^{\beta \gamma} \widehat\nabla^{\alpha} \lambda^I \widehat\nabla_{\gamma} \widehat\nabla_{\beta}\widehat \nabla_{\alpha} \lambda^J\cr
&\left.+\frac{5}{6} \square R^{\mu \nu} \lambda^I \widehat\nabla_{\mu} \widehat\nabla_{\nu} \lambda^J+\frac{22}{3} R^{\mu \nu} \widehat\nabla_{\mu} \widehat\nabla_{\nu} \lambda^I \widehat\square \lambda^J-\frac{5}{3} \nabla^{\mu} R^{\alpha \beta} \nabla_{\mu} R_{\alpha \beta} \lambda^I\lambda^I\right\}+\mathcal{O}\left(\lambda^{4}\right)\;,
\end{align}
where $c$ is the central charge of the system. From this expression it is apparent that the $\alpha$ parameter entering \eqref{eq:12} is the normalisation of $\langle T \OO \OO\rangle$  which, in the unbroken phase, can be related to the normalization of $\langle \OO \OO\rangle$. However, there is no information about $\beta$, since the part of $\delta_\sigma \gamma_{\mu\nu}$ that it parametrises does not contribute to $\delta_\sigma(C_{\mu \nu \rho\sigma} C^{\mu \nu \rho\sigma} )$.\footnote{Note that when computing $\delta_\sigma(C_{\mu \nu \rho\sigma} C^{\mu \nu \rho\sigma} )$, all the $\nabla_\mu$ operators hitting $\delta_\sigma \gamma_{\mu\nu}$ can be promoted to their hatted versions, i.e. $\nabla_\mu\delta_\sigma \gamma_{\nu\rho}=\widehat\nabla_\mu\delta_\sigma \gamma_{\nu\rho}$.} The WZ consistency condition for the anomaly \eqref{eq:14} is satisfied up to terms that vanish when $\nabla_i G_{IJ}=0$ and up to bundle-curvature terms ($F$-terms), since in their absence our expression then reverts to the one of \cite{Schwimmer:2019efk}.\footnote{By $F$-terms we denote contributions that vanish when $F_{ij}=0$, but do not vanish when $\nabla_i G_{IJ}=0$.} Our next goal will be to introduce new terms $\mathcal{A}^F$ to the anomaly \eqref{eq:14} that remove the $F$-terms in the WZ consistency condition.

For the purposes of this paper, it will be enough to determine the new terms that are needed to make $\mathcal{A}+\mathcal{A}^F$ WZ consistent to leading order around flat spacetime, $\gamma_{\mu\nu} \simeq \delta_{\mu\nu} + \ldots$. We will therefore ignore in $\mathcal{A}$, $\mathcal{A}^F$ terms quadratic (or higher) in the spacetime curvature, like the Weyl-tensor squared. However, terms linear in the spacetime curvature must be taken into account, as the flat spacetime limit of $\delta_\sigma R_{\mu\nu\rho\sigma}$ does not vanish, c.f.\ \eqref{eq:23}. Accordingly, we will work with the classical Weyl variation of the spacetime metric and up to quadratic order in the $\lambda$s. In summary, we want to identify the terms $\mathcal{A}^F_{\rm flat}$ that can remove all $F$-terms from the flat-spacetime limit of the WZ condition for $\mathcal{A}_{\rm flat}$, which reads
\begin{align}
  \label{eq:15}
\mathcal{A}_{\rm flat}\propto & \,  G_{IJ}\Big[\widehat\square \lambda^I \widehat\square^{2} \lambda^J -\frac{1}{3} R \widehat\square \widehat\nabla^{\mu} \lambda^I \widehat\nabla_{\mu} \lambda^J+\frac{1}{36} \square R \lambda^I \widehat\square \lambda^J-\frac{16}{9} R \widehat\square \lambda^I \widehat\square \lambda^J\cr
&+\nabla^{\mu} R \widehat\nabla_{\mu} \lambda^I \widehat\square \lambda^J+\frac{1}{6} R \lambda^I \widehat\square^{2} \lambda^J-4 R^{\mu \nu} \widehat\nabla_{\mu} \lambda^I \widehat\square \widehat\nabla_{\nu} \lambda^J-\frac{5}{18} \nabla^{\mu} \nabla^{\nu} R \lambda^I \widehat\nabla_{\mu} \widehat\nabla_{\nu} \lambda^J\cr
&-\frac{5}{9} R \widehat\nabla_{\mu} \widehat\nabla_{\nu} \lambda^I \widehat\nabla^{\mu} \widehat\nabla^{\nu} \lambda^J+\frac{10}{3} R^{\beta \gamma} \widehat\nabla^{\alpha} \lambda^I \widehat\nabla_{\gamma} \widehat\nabla_{\beta} \widehat\nabla_{\alpha} \lambda^J+\frac{5}{6} \square R^{\mu \nu} \lambda^I \widehat\nabla_{\mu} \widehat\nabla_{\nu} \lambda^J\cr
  &+\frac{22}{3} R^{\mu \nu} \widehat\nabla_{\mu} \widehat\nabla_{\nu} \lambda^I \widehat\square \lambda^J\Big].
\end{align}

The $F$-terms that enter the flat spacetime limit of the WZ consistency condition for $\mathcal{A}_{\rm flat}$ are\footnote{To simplify our expressions, we will denote $F_{\mu\nu Q}^P:=(F_{\mu\nu})^P_{\;\,Q}$, $F^{\mu\nu P}_Q:=(F^{\mu\nu})^P_{\;\,Q}$ and $F^{\mu P}_{\nu Q}:=(F_\nu^{\;\,\mu})^P_{\;\,Q}$. Analogous definitions will apply to $F_{ij}$.}
\begin{align}
\label{toBecancelled}
\int d^4x& \sqrt{\gamma}\delta\sigma_{[1} \nabla^\mu \delta\sigma_{2]}G_{IJ}\times\cr
&\Bigg[-8F_{\nu  \rho L}^{J}F^{\nu  \rho I}_{K}\lambda^K \widehat\nabla_\mu  \lambda^L-\frac{40}3F_{\mu K}^{\rho I}F_{\nu  \rho L}^J\lambda^K \widehat\nabla^\nu  \lambda^J-\frac{20}3\lambda^K\widehat{\square}\lambda^L\widehat \nabla F_{\mu K}^{\rho I}\cr
&+\frac43 \wn_\nu  \lambda^L \wn^\nu   \lambda^J \wn F_{\mu L}^{\rho I}+\frac{40}3\lambda^K\wn^\nu  \wn_\mu \lambda^J\wn F_{\nu  K}^{\rho  I}+\frac{20}{3}F_{\mu K}^{\nu  I}\lambda^K\lambda^L \wn_\rho   F_{\nu  L}^{\rho J}\cr
&+8\lambda^K\wn^\nu  \lambda^J \ws F_{\mu \nu  K}^I-28F_{\mu\nu  K}^I \wn \lambda^K \ws\lambda^J+\frac43 F_{\mu \nu  K}^I\lambda^K \ws \wn^\nu   \lambda^J\cr
&+16 \wn \lambda^J \wn_\rho  F_{\mu \nu  L}^I\wn^\rho  \lambda^L-\frac{32}3F_{\nu  \rho K}^I\wn^\nu   \lambda^K \wn^\rho \wn_\mu \lambda^J-\frac83F_{\mu \rho K}^I\wn^\nu  \lambda^K\wn^\rho \wn_\nu  \lambda^J\Bigg]\;.
\end{align}
One arrives at this expression by making use of the Bianchi identity for $F_{\mu\nu}$, and rearranging the order of the  $\wn$-operators into terms of the type $\wn^{(n+2)}G_{IJ}$.\footnote{For example, one can rewrite expressions of the type $G_{K (I}\wn^{(n)}(F_{\mu\nu})^K_{\;J)}$ solely in terms of $G_{K I}\wn^{(m)}(F_{\mu\nu})^K_{\;J}$ with $m\le n-2$, and terms that vanish when $\nabla_i G_{IJ}=0$. In particular, for $n=0$ we have that $F_{\mu\nu (IJ)}=0$ when $\nabla_i G_{IJ}=0$, with $F_{\mu\nu IJ}:=G_{KI}(F_{\mu\nu})^K_{\;J}$.}

To cancel the terms in \eqref{toBecancelled}, we start with the most general linear ansatz that is quadratic in the sources $\lambda$, that vanishes when $F_{ij}=0$, and that is at most linear in the spacetime curvature. Moreover, since \eqref{toBecancelled} involves only $F_{\mu\nu}=\lambda^i_\mu\lambda^j_\nu F_{ij}$, each $F_{ij}$ must come contracted with corresponding factors of $\lambda^i_\mu$. Without any additional algebraic simplifications, nor through identifying redundancies due to Weyl-cohomologically trivial terms, we have determined using the xAct Mathematica package \cite{Martin-Garcia:2007bqa,Martin-Garcia:2008yei,Mart_n_Garc_a_2008,Brizuela:2008ra,Nutma:2013zea,Frob:2020gdh} that such an ansatz comprises $\sim 1500$ terms. These contribute to the WZ consistency condition with two classes of terms: those that can be rewritten solely in terms of the combination $F_{\mu\nu}= \lambda^i_\mu \lambda^j_\nu F_{ij}$ and those where the curvature components $F_{ij}$ of the $\lambda-$bundle necessarily appear explicitly. We require that the former cancel out the terms in \eqref{toBecancelled} and the latter cancel out by themselves. This yields a solution that fixes some of the coefficients of the linear ansatz and leaves the remaining undetermined. By setting the undetermined coefficients to zero the resulting expression has the following 126 terms:\footnote{Our Mathematica notebook with the full solution can be made available upon request.}  
{\small
\begin{align}
&\mathcal{A}_{\rm flat}^F=G_{IJ}\Bigg[-\frac{19}{6}F_{\nu  \rho  L }^I  F^{\nu  \rho  J }_K  \m^K  \ws \m^L +\frac {359}{144}F_{ i    j  K }^I  R  \m^{ i  \mu}\m^J \m^K \wn_\mu \wn_\nu \m^{  j  \nu }-\frac{23}3 F^{\nu  \rho  I }_K  \m^K  \nabla  _\rho  R_{\mu \nu }\wn^\mu  \m^J \cr
&-\frac {35}{9}F_{\mu  K }^{\nu  I }\m^J  \nabla  _\rho  R^\rho _\nu \wn^\mu  \m^K  +\frac {49}6F_{\nu  \rho  L }^I  F^{\nu  \rho  J }_K  \wn_\mu  \m^L \wn^\mu  \m^K +\frac {151}{24}\m^{ i \mu }\m^J \m^K  \wn_\mu \wn^\nu \wn_\rho  \m^{  j  \rho }\wn_\nu  F_{ i    j  K }^I \cr
&-\frac 13 F_{ i    j  K }^I  R \m^{ i \mu } \m^K  \wn_\mu  \m^J  \wn_\nu  \m^{  j  \nu }+F_{ i    j  K }^I  R \m^{ i \mu }\m^J  \wn_\mu  \m^K  \wn_\nu  \m^{  j  \nu }-\frac {359}{144}F_{ i    j  K }^I  R \m^{ i \mu }\m^J  \m^K  \ws \m^{  j }_\mu \cr
&+\frac {89}{24}\m^K  \wn^\mu  \m^J  \wn_\nu \ws F_{\mu  K }^{ \nu I }+\frac {95}{12}\m^J \wn^\mu \m^K \wn_\nu \ws F_{\mu  K }^{ \nu I }+\frac {215}8\m^{ i \mu }\m^J \m^K \wn_{\mu }\wn_\rho  F_{ i    j  K }^I \wn_\nu \wn^\rho \m^{  j  \nu }\cr
&+\frac {255}{4}F_{ i    j  K }^I  \m^J \m^K  \wn_\nu \wn_\mu \wn_\rho  \m^{  j  \rho } \wn^\nu \m^{ i \mu }+\frac {395}{6}\m^J \m^K  \wn_\mu  F_{ i    j  K }^I  \wn_\nu \wn_\rho  \m^{  j  \rho }\wn^\nu  \m^{ i \mu }\cr
&+\frac {2519}{96}F_{ i    j  K }^I  \m^K  \wn_\mu  \m^J  \wn_\nu  \wn_\rho  \m^{  j  \rho }\wn^\nu \m^{ i \mu }+\frac {2509}{96}F_{ i    j  K }^I  \m^J  \wn_\mu  \m^K  \wn_\nu \wn_\rho  \m^{  j  \rho } \wn^\nu  \m^{ i \mu }\cr
&-\frac {1139}{24}F_{ i    j  K }^I  \m^J  \m^K \wn_\nu \ws \m^{  j }_\mu \wn^\nu \m^{ i \mu }+\frac 23 F_{ i    j  K }^I  \m^{ i \mu } \m^J  \m^K  \nabla  _\rho  R^\rho _\nu  \wn^\nu \m^  j _\mu+\frac {35}8F_{ i    j  K}^I  \m^{ i  \mu}\m^K \wn_\mu\ws\m^  j _\nu \wn^\nu \m^J \cr
&+\frac {475}8\m^{ i  \mu}\m^K  \wn_\mu\wn_\rho  \m^{  j  \rho }\wn_\nu  F_{ i    j  K}^I \wn^\nu  \m^J +\frac {2519}{96}F_{ i    j  K }^I  \m^{ i  \mu}\m^K \wn_\nu \wn_\mu\wn_\rho \m^{  j  \rho }\wn^\nu \m^J \cr
&-\frac {105}{8}\m^{ i  \mu}\m^K \wn_\mu F_{ i    j  K }^I \wn_\nu \wn_\rho \m^{  j  \rho }\wn^\nu \m^J -\frac {2039}{96}F_{ i    j  K }^I  \m^{ i  \mu}\m^K \wn_\nu \ws \m^  j _\mu\wn^\nu \m^J \cr
&+\frac {55}{12}F_{ i    j  K }^I \m^{ i  \mu}\m^J \wn_\mu  \ws \m^  j _\nu \wn^\nu \m^K +\frac {2509}{96}F_{ i    j  K }^I \m^{ i  \mu}\m^J  \wn_\nu \wn_\mu \wn_\rho  \m^{  j  \rho }\wn^\nu \m^K \cr
&+\frac 94\m^{ i  \mu}\m^J \wn_\mu F_{ i    j  K }^I \wn_\nu \wn_\rho \m^{  j  \rho }\wn^\nu  \m^K -\frac {2029}{96}F_{ i    j  K }^I  \m^{ i  \mu}\m^J \wn_\nu \ws \m^  j _\mu\wn^\nu \m^K +11F_{\mu L }^{\rho  J }F_{\nu  \rho  K }^I \wn^\mu \m^K  \wn^\nu \m^L \cr
&-F_{\mu K }^{\rho  J }F_{\nu  \rho   L }^ I  \wn^\mu \m^K  \wn^\nu  \m^L +\frac {473}{8}\m^{ i \mu }\m^J  \m^K  \wn_\nu  F_{ i    j  K }^I  \wn^\nu \wn_\mu\wn_\rho  \m^{  j  \rho }-\frac 23 F_{\nu  \rho  K }^I  \wn_\mu \wn^\rho  \m^K  \wn^\nu  \wn^\mu \m^J \cr
&+\frac {40}3F_{\mu K }^{\rho  J }F_{\nu  \rho  L }^I  \m^K  \wn^\nu \wn^\mu \m^L +\frac {241}{24}\m^J \m^L \wn_\mu\m^{ i  \mu}\wn_\nu  F_{ i    j  K }^I  \wn^\nu \wn_\rho  \m^{  j  \rho }-\frac {89}{2}\m^{ i \mu }\m^J \m^K  \wn_\nu  F_{ i    j  K }^I  \wn^\nu  \ws \m^  j _\mu\cr
&+\frac {623}{24}\m^J \m^K \wn_\nu \wn^\rho \m^  j _\mu\wn^\nu \m^{ i  \mu}\wn_\rho  F_{ i    j  K }^I +\frac {73}8\m^{ i  \mu}\m^K  \wn_\mu \wn^\rho \m^  j _\nu \wn^\nu \m^J \wn_\rho  F_{ i    j  K }^I \cr
&+\frac {175}{12}\m^{ i  \mu}\m^K \wn_\nu \wn^\rho \m^  j _\mu\wn^\nu \m^J \wn_\rho  F_{ i    j  K }^I +\frac {22}{3} \m^{ i  \mu}\m^J \wn_\mu\wn^\rho \m^{  j }_\nu \wn^\nu \m^K  \wn_\rho  F_{ i    j  K }^I \cr
&+\frac {14}3R^{\nu \rho }\m^K \wn^\mu \m^J \wn_\rho F_{\mu \nu K }^I +\frac {83}{24}R^\nu _\mu \m^K  \m^K \wn^\mu  \m^J \wn_\rho F_{\nu  K }^{\rho  I }+2 F_{ i    j  K }^I  \m^{ i  \mu}\m^K  \wn_\mu\ws \m^J  \wn_\rho  \m^{  j  \rho }\cr
&+2F_{ i    j  K }^I \m^{ i  \mu}\wn_\mu \m^K  \ws \m^J  \wn_\rho \m^{  j  \rho }-\frac {85}6 \m^K  \wn_\mu \m^J \wn_\nu  F_{ i    j  K }^I  \wn^\nu \m^{ i  \mu} \wn_\rho  \m^{  j  \rho }\cr
&+2 \m^J \wn_\mu \m^K  \wn_\nu  F_{ i    j  K }^I  \wn^\nu \m^{ i  \mu} \wn_\rho  \m^{  j  \rho }+\frac {77}{24}F_{ i    j  K }^I  R_{\mu \nu }\m^{ i  \mu} \m^K  \wn^\nu  \m^J \wn_\rho  \m^{  j  \rho }\cr
&+26 \m^{ i  \mu} \m^K  \wn_\nu \wn_\mu F_{ i    j  K }^I  \wn^\nu  \m^J  \wn_\rho  \m^{  j  \rho }-4F_{ i    j  K }^I  R_{\mu \nu } \m^{ i  \mu} \m^J  \wn^\nu \m^K  \wn_\rho  \m^{  j  \rho }\cr
&-8F_{ i    j  K }^I  \m^{ i  \mu} \wn_\mu\wn_\nu \m^J  \wn^\nu \m^K  \wn_\rho \m^{  j  \rho }-\frac {247}{12}\m^{ i  \mu} \m^J \wn_\nu \wn_\mu F_{ i    j  K }^I  \wn^\nu  \m^K  \wn_\rho  \m^{  j  \rho }\cr
&-\frac {295}{12}\m^{ i  \mu} \m^J \m^K \wn_\nu \wn^\rho  \m^{  j  \nu } \wn_\rho \wn_\mu F_{ i    j  K }^I +\frac {215}{24}\m^{ i  \mu} \m^J \m^K  \wn_\nu  \m^{  j  \nu }\wn_\rho \wn_\mu\wn^\rho   F_{ i    j  K }^I \cr
&+\frac {523}{24}\m^{ i  \mu} \m^J \m^K  \wn_\nu  F_{ i    j  K }^I \wn_\rho \wn_\mu \wn^\rho \m^{  j  \nu }+\frac {151}{24}F_{ i    j  K }^I  \m^{ i  \mu}\m^J \m^K  \wn_\rho  \wn_\mu \wn^\rho \wn_\nu \m^{  j  \nu }\cr
&+\frac {215}{24}\m^{ i  \mu}\m^J \m^K \wn_\mu\wn^\rho  \m^{  j  \nu }\wn_\rho \wn_\nu  F_{ i    j  K }^I -\frac {337}{24}\m^{ i  \mu}\m^J \m^K \wn^\nu \m^  j _\mu\wn_\rho \wn_\nu \wn^\rho  F_{ i    j  K }^I \cr
&-\frac {11}2 \m^{ i  \mu}\m^J \m^K  \wn_\mu F_{ i    j  K }^I  \wn_\rho \wn_\nu \wn^\rho  \m^{  j  \nu } -\frac {151}{24}F_{ i   j  K }^I  \m^J \m^K  \wn_\mu \m^{ i  \mu} \wn_\rho \wn_\nu \wn^\rho  \m^{  j  \nu }\cr
& -\frac {35}{8}F_{ i    j  K }^I  \m^{ i  \mu}\m^K  \wn_\mu \m^J  \wn_\rho \wn_\nu \wn^\rho  \m^{  j  \nu }-\frac {55}{12} F_{ i    j  K }^I  \m^{ i  \mu}\m^J \wn_\mu \m^K  \wn_\rho \wn_\nu \wn^\rho  \m^{  j  \nu }\cr
&+\frac {215}{24}\m^J  \m^K  \wn_\rho \wn_\nu \wn^\rho \wn_\mu F^{\mu \nu I }_K  -\frac {151}{24} F_{ i    j  K }^I  \m^{ i  \mu} \m^J  \m^K  \wn_\rho \wn_\nu \wn^\rho  \wn^\nu  \m^  j _\mu\cr
&-\frac {215}{24}\m^{ i  \mu} \m^J  \m^K  \wn_\nu  F_{ i    j  K }^I  \wn_\rho  \wn^\nu  \wn^\rho  \m^  j _\mu+\frac {215}{8} \m^{ i  \mu}\m^J  \m^K  \wn_\mu \wn_\nu  \m^{  j  \nu } \ws F_{ i    j  K }^I \cr
&-\frac {47}{2}\m^{ i  \mu}\m^K  \wn_\mu \m^J  \wn_\nu \m^{  j  \nu } \ws F_{ i    j  K }^I +\frac {121}{12} \m^{ i  \mu}\m^J  \wn_\mu \m^K  \wn^\nu  \m^{  j  \nu } \ws F_{ i    j  K }^I \cr
&-\frac {215}{8}\m^{ i  \mu} \m^J \m^K  \ws \m^  j _\mu \ws F_{ i    j  K }^I +\frac {1157}{48}\m^{ i  \mu} \m^K  \wn_\nu \m^J \wn^\nu \m^  j _\mu\ws F_{ i    j  K }^I \cr
&-\frac {455}{48} \m^{ i  \mu}\m^J  \wn_\nu \m^K  \wn^\nu \m^  j _\mu \ws F_{ i    j  K }^I  +\frac {35}{8} F_{ i    j  K }^I  \m^K  \wn^\mu \m^J  \wn_\nu  \m^{ i  \nu } \ws \m^  j _\mu \cr
&+\frac {55}{12} F_{ i    j  K }^I  \m^J \wn^\mu \m^K  \wn_\nu  \m^{ i  \nu } \ws \m^  j _\mu -\frac {3073}{48}\m^{ i  \mu}\m^K  \wn_\nu  F_{ i    j  K }^I  \wn^\nu  \m^J  \ws \m^  j _\mu \cr
&+\frac {431}{16}\m^{ i  \mu} \m^J  \wn_\nu  F_{ i    j  K }^I  \wn^\nu  \m^K  \ws \m^  j _\mu+\frac {215}{8} F_{ i    j  K }^I  \m^J \m^K  \wn_\mu \wn^\nu \m^{ i  \mu} \ws \m^  j _\nu  \cr
&-\frac {123}2 \m^J \m^K  \wn_\mu F_{ i    j  K }^I \wn^\nu \m^{ i  \mu} \ws \m^{  j }_\nu -\frac {3299}{96} F_{ i    j  K }^I  \m^K  \wn_\mu \m^J  \wn^\nu  \m^{ i  \mu} \ws \m^{  j }_\nu \cr
&-\frac {3349}{96}F_{ i    j  K }^I \m^J  \wn_\mu \m^K  \wn^\nu  \m^{ i  \mu} \ws \m^{  j }_\nu +\frac {163}{12} \m^{ i  \mu}\m^K  \wn_\mu F_{ i    j  K }^I  \wn^\nu  \m^J  \ws \m^  j _\nu \cr
&+\frac {161}{12}\m^{ i  \mu}\m^J \wn_\mu F_{ i    j  K }^I  \wn^\nu  \m^K  \ws \m^{  j }_\nu -\frac {229}6 F_{ i    j  K }^I  \m^J  \m^K  \wn^\nu  \wn_\mu \m^{ i  \mu} \ws \m^{  j  }_\nu \cr
&-\frac {10}3\m^J  \m^K  \wn_\mu \m^{ i  \mu} \wn_\nu  F_{ i    j  K }^I  \ws \m^{  j  \nu }-\frac {161}{24} \m^{ i  \mu} \m^K  \wn_\mu \m^J  \wn_\nu  F_{ i    j  K }^I  \ws \m^{  j  \nu }\cr
&-\frac {59}{12}\m^{ i  \mu}\m^J  \wn_\mu \m^K  \wn_\nu  F_{ i   j  K }^I  \ws \m^{  j  \nu }+\frac {65}{24} \m^{ i  \mu} \m^J  \m^K  \wn_\nu  \m^{  j  \nu } \ws \wn_\mu F_{ i    j  K }^I \cr
&+\frac {307}{12}F_{ i    j  K }^I  \m^{ i  \mu}\m^J \m^K  \ws \wn_\mu\wn_\nu \m^{  j  \nu }+\frac {239}{12} \m^{ i  \mu} \m^J  \m^K  \wn^\nu  \m^  j _\mu \ws \wn_\nu  F_{ i    j  K }^I \cr
&-\frac {151}{24}F_{ i    j  K }^I  \m^J \m^K  \wn^\nu  \m^{ i  \mu} \ws \wn_\nu  \m^  j _\mu+\frac {57}8\m^{ i  \mu}\m^J \m^K  \wn_\mu F_{ i    j  K }^I  \ws \wn_\nu \m^{  j  \nu } \cr
&+\frac {151}{24}F_{ i    j  K }^I  \m^J  \m^K  \wn_\mu \m^{ i  \mu} \ws \wn_\nu \m^{  j  \nu }+\frac {215}{12} \m^J  \m^K  \ws \wn_\nu \wn_\mu F^{\mu \nu I }_K \cr
&-\frac {247}{12} F_{ i    j  K }^I  \m^{ i  \mu} \m^J  \m^K  \ws \ws \m^  j _\mu -\frac {22}3 \m^{ i  \mu} \m^J  \m^K  \wn_\nu  F_{ i    j  K }^I  \ws \wn^\nu  \m^  j  _\mu \cr
&+6 F_{\mu \nu K }^I  \wn^\mu \m^K  \ws \wn^\nu  \m^J  +\frac {35}4 F_{ i    j  K }^I  \m^K  \wn^\mu \m^J  \wn_\rho  \wn_\nu  \m^  j _\mu \wn^\rho  \m^{ i  \nu }\cr
&+\frac {55}6 F_{ i    j  K }^I  \m^J  \wn^\mu \m^K  \wn_\rho  \wn_\nu  \m^  j _\mu \wn^\rho  \m^{ i  \nu }+\frac {215}{24} \m^J \m^K  \wn_\nu \wn_\rho  F_{ i    j  K }^I \wn^\nu  \m^{ i  \mu} \wn^\rho  \m^  j _\mu \cr
&+\frac {19}{24} F_{ i    j  K }^I  R_{\nu  \rho }\m^{ i  \mu} \m^K  \wn^\nu \m^J  \wn^\rho  \m^  j _\mu -\frac {215}{24} \m^{ i  \mu} \m^K  \wn_\nu  \wn_\rho  F_{ i    j  K }^I  \wn^\nu  \m^J  \wn^\rho  \m^  j _\mu \cr
&+\frac {449}{12} \m^{ i  \mu} \m^J  \wn_\nu  \wn_\rho  F_{ i    j  K }^I  \wn^\nu  \m^K  \wn^\rho  \m^  j _\mu -\frac {471}8 \m^J  \m^K  \wn^\nu  \m^{ i  \mu} \wn_\rho  \wn_\nu  F_{ i    j  K }^I  \wn^\rho  \m^  j _\mu \cr
&-\frac {875}{48} \m^K  \wn_\mu \m^J  \wn^\nu  \m^{ i  \mu} \wn_\rho  F_{ i    j  K }^I  \wn^\rho  \m^  j  _\nu -\frac {809}{48} \m^J  \wn_\mu \m^K  \wn^\nu  \m^{ i  \mu} \wn_\rho  F_{ i    j  K }^I  \wn^\rho  \m^  j _\nu  \cr
&-2 F_{ i    j  K }^I  R_{\nu  \rho } \m^{ i  \mu} \m^K  \wn_\mu \m^J  \wn^\rho  \m^{  j  \nu }+\frac {215}{24} \m^{ i  \mu} \m^J  \m^K  \wn_\mu \wn_\rho  \wn_\nu  F_{ i    j  K }^I  \wn^\rho  \m^{  j  \nu }\cr
&+2 F_{ i    j  K }^I  R_{\mu \rho } \m^{ i  \mu} \m^K  \wn_\nu  \m^J  \wn^\rho \m^{  j  \nu }+\frac {391}{48} \m^{ i  \mu} \m^K  \wn_\mu \wn_\rho  F_{ i    j  K }^I  \wn_\nu  \m^J  \wn^\rho  \m^{  j  \nu }\cr
&+\frac {137}{16} \m^{ i  \mu} \m^I   \wn_\mu \wn_\rho  F_{ i    j  K }^J \wn_\nu  \m^K  \wn^\rho  \m^{  j  \nu }+\frac {655}{48} \m^K  \wn_\mu \m^J  \wn_\nu  \m^{ i  \mu} \wn_\rho  F_{ i    j  K }^I  \wn^\rho  \m^{  j  \nu }\cr
&-\frac {65}{48} \m^J  \wn_\mu \m^K  \wn_\nu  \m^{ i  \mu} \wn_\rho  F_{ i    j  K }^I  \wn^\rho  \m^{  j  \nu }-\frac {397}{48} \m^{ i  \mu} \m^K  \wn_\nu  \m^J  \wn_\rho  \wn_\mu F_{ i    j  K }^I  \wn^\rho  \m^{  j  \nu }\cr
&-\frac {377}{48} \m^{ i  \mu} \m^J  \wn_\nu  \m^K  \wn_\rho  \wn_\mu F_{ i    j  K }^I  \wn^\rho  \m^{  j  \nu }+\frac {153}{16} \m^{ i  \mu} \m^J  \m^K  \wn_\rho  \wn_\mu \wn_\nu  F_{ i    j  K }^I  \wn^\rho  \m^{  j  \nu }\cr
&-\frac {17}3 \m^J  \m^K  \wn_\mu \m^{ i  \mu} \wn_\rho  \wn_\nu  F_{ i    j  K }^I  \wn^\rho  \m^{  j  \nu }-\frac {29}{24} \m^{ i  \mu} \m^K  \wn_\mu \m^J  \wn_\rho  \wn_\nu  F_{ i    j  K }^I  \wn^\rho  \m^{  j  \nu }\cr
&-\frac {29}{24}\m^{ i  \mu}\m^J  \wn_\mu \m^K  \wn_\rho  \wn_\nu  F_{ i    j  K }^I  \wn^\rho  \m^{  j  \nu }-\frac {311}{16} \m^{ i  \mu} \m^J  \m^K  \wn_\rho  \wn_\nu  \wn_\mu F_{ i    j  K }^I  \wn^\rho  \m^{  j  \nu }\cr
&-\frac {215}{24} \m^J  \m^K \wn^\nu \m^{i \mu}  \wn_\rho  F_{ i    j  K }^I  \wn^\rho  \wn_\nu  \m^  j _\mu +\frac {31}8 \m^{ i  \mu} \m^J  \m^K  \wn_\mu \wn_\rho  F_{ i    j  K }^I  \wn^\rho  \wn_\nu  \m^{  j  \nu }\cr
&+\frac {83}{12}\m^{ i  \mu} \m^J  \m^K  \wn_\rho  \wn_\mu F_{ i    j  K }^I  \wn^\rho  \wn_\nu  \m^{  j  \nu }-2 F_{\mu \rho  K }^I  \wn^\mu \m^K  \wn^\rho  \ws \m^J \cr
&+\frac {151}{24} F_{ i    j  K }^I  \m^J  \m^K  \wn_\nu  \wn_\rho  \m^  j _\mu \wn^\rho  \wn^\nu  \m^{ i  \mu}-\frac {215}{24} \m^{ i  \mu} \m^J  \m^K  \wn_\nu  \wn_\rho  F_{ i    j  K }^I  \wn^\rho  \wn^\nu  \m^  j _\mu \cr
&+\frac {497}{24}\m^{ i  \mu} \m^J  \m^K  \wn_\rho  \wn_\nu  F_{ i    j  K }^I  \wn^\rho  \wn^\nu  \m^  j _\mu+12F_{\nu  \rho  K }^I  \wn^\mu \m^K  \wn^\rho \wn^\nu  \wn_\mu \m^J \cr
&-\frac {71}{16}R_{\mu \sigma \nu \rho } \m^K  \wn^\mu \m^J  \wn^\sigma F^{\nu  \rho  I }_K  -\frac {41}8 F_{ i    j  K }^I  R_{\mu \rho  \sigma \nu } \m^{ i  \mu}\m^K  \wn^\nu  \m^J  \wn^\sigma \m^{  j  \rho }\cr
&-\frac {22}3 F_{ i   j  K }^I  R_{\mu \rho  \nu \sigma} \m^{ i  \mu} \m^J  \wn^\nu  \m^K  \wn^\sigma \m^{  j  \rho }\Bigg]\;.
\label{counterTerms126}
\end{align}}
In particular, one can prove using Mathematica that it is impossible to cancel out the contributions in \eqref{toBecancelled} using new terms that contain exclusively the combination $F_{\mu\nu}$.

The flat-spacetime limit is sufficient for the purposes of this paper. Nevertheless, it is interesting to ask how \eqref{counterTerms126} would be modified in the case of arbitrary spacetime curvature. In such a case, one should also add to the ansatz terms that are at least quadratic in the Riemann tensor and linear in the vector-bundle curvature. Dimensional analysis suggests that the only possibility is of the type $G_{RP}F_{\mu\nu\;Q}^R R_{abcd} R_{efgh} \m^P \m^Q$, however by taking into account all possible spacetime contractions all such terms vanish. Consequently, our original ansatz should be sufficient towards determining the anomaly functional for any curved (spacetime and vector-bundle) background. This is a well-defined but computationally challenging problem, to which we hope to return in the future.

\subsection{$\Delta=2$ Operators}\label{WZdelta2}

We have left the case of $\Delta=2$ operators for last as it is trivial. The anomaly is encoded in
\beq
\label{wz4aa}
\delta_\sigma W \propto \int d^4 x \sqrt{\gamma}\, \delta \sigma \, G_{IJ} \lambda^I  \lambda^J
~. 
\eeq
The above automatically satisfies the WZ consistency condition and does not involve the connection $A$. Hence, one cannot infer anything about $\nabla_i G_{I J}$ from this expression. In the next section, we will return to the case of $\Delta=2$ type-B conformal anomalies in the context of 4D $\NN=2$ SCFTs, where supersymmetry will allow us to say more.

\section{$\Delta=2$ CBOs in 4D $\NN=2$ SCFTs}\label{delta2}

In this section we will focus on CBOs $\OO_I$ (and their complex conjugates $\bar \OO_I$) with scaling dimension $\Delta=2$ in 4D $\NN=2$ SCFTs. We will argue using Poincaré supersymmetry that the $\Delta=2$ type-B Weyl anomalies are the same as the type-B Weyl anomalies of the exactly marginal $\Delta=4$ operators. This is obvious in the conformally symmetric phase (see e.g. \cite{Niarchos:2019onf}), but requires a less straightforward argument in phases with spontaneously broken conformal symmetry. We will outline the argument in Sec~\ref{arg} and provide tree-level supporting evidence for its validity in Sec~\ref{check}. Once the relation with the exactly marginal Weyl anomalies is established, the result $\nabla G=0$ for $\Delta=2$ anomalies follows from Eq.~\eqref{eq:marginal}.

\subsection{Anomalies Related by Poincar\'e Supersymmetry}
\label{arg}

The exactly marginal operators of the $\mathcal N=2$ SCFT are of the form\footnote{We use shorthand notation to denote the usual adjoint action of the supersymmetry generators.}
\beq
\label{cb2aa}
\Phi_i \propto Q^4 \cdot \OO_I \delta{_i^I}~, ~~ \bar \Phi_i \propto \bar Q^4 \cdot \bar \OO_I \delta{_i^I}
~.
\eeq
In the conformal phase it is straightforward to relate the anomaly of the $\Delta=2$ operators $\OO_I$ to the anomaly of the exactly marginal operators $\Phi_i$ by looking at the corresponding 2-point functions \eqref{extractingCFT}. The Ward identity for Poincaré supercharges
\begin{align}
  \label{eq:20}
  \sum_{k=1}^{n}\left\langle\varphi_{1}\left(x_{1}\right) \ldots Q \cdot \varphi_{k}\left(x_{k}\right) \ldots \varphi_{n}\left(x_{n}\right)\right\rangle=0\;,
\end{align}
can be used to move the supercharges around so as to arrive at \cite{Papadodimas:2009eu}
\begin{align}
  \label{eq:2}
  \left\langle\mathcal{O}_{I}(x_1) \bar{\mathcal{O}}_{J}(x_2)\right\rangle \propto \Box_{x_2}^2\left\langle\Phi_{i}(x_2) \bar{\Phi}_{j}(x_2)\right\rangle \delta_I^i \delta_J^j\;,
\end{align}
where the constant of proportionality depends on conventions and will be fixed momentarily. 

In a general phase, the type-B anomaly of interest is captured by a particular contact term in the 3-point function \eqref{extractingHB}
\beq
\label{cb2ac}
\langle T(x) \OO_I(x_1) \bar \OO_J (x_2) \rangle
~,
\eeq
where $T\equiv T^\mu_\mu$ is the trace of the energy-momentum tensor. The energy-momentum tensor of the $\NN=2$ SCFT belongs to a superconformal multiplet with a scalar superconformal primary $\TT$ that obeys the shortening conditions $(Q^\II)^2 \cdot\TT =0$, $(\bar Q_\II)^2 \cdot\TT=0$ (for $\II=1,2$ the $SU(2)_R$ R-symmetry index), and is of the form (suppressing spacetime indices, spinor indices and sigma-matrices on the RHS)
\beq 
\label{cb2ab}
T_{\mu\nu} = Q^1 \cdot Q^2 \cdot\bar Q_1 \cdot \bar Q_2 \cdot\TT + c_1 Q^1 \cdot\bar Q_1 \cdot\d \TT + c_2 Q^2 \cdot\bar Q_2 \cdot\d \TT + c_3 \d^2 \TT
~.
\eeq

In phases with spontaneously broken conformal symmetry it is less straightforward to relate \eqref{cb2ac} to $\langle T(x) \Phi_i(x_1) \bar \Phi_j (x_2) \rangle$ by applying Ward identities. In vacua, where Poincar\'e supersymmetry is unbroken, as e.g.\ on the Coulomb or Higgs branch of $\NN=2$ SCFTs, one can still use the integrated form of the Ward identities \eqref{eq:20}, but their application on 3-point functions of the form \eqref{cb2ac} is complicated. However, since we only care about a contact term in the limit of vanishing momentum for the energy-momentum tensor, it may be natural to anticipate that terms in $T_{\mu\nu}$ with explicit spacetime derivatives (like the $c_1,c_2,c_3$ terms in \eqref{cb2ab}) will not contribute to the anomaly. Assuming such terms can be dropped, in the 3-point function 
\beq
\label{cb2ae}
\langle T(x) (Q^4\cdot\OO_I)(x_1) (\bar Q^4 \cdot \bar \OO_J) (x_2) \rangle
\eeq
with two exactly marginal operators, only 
\beq
\label{cb2af}
\langle (Q^1\cdot Q^2 \cdot \bar Q_1 \cdot \bar Q_2\cdot  \TT)(x) (Q^4\cdot \OO_I)(x_1) (\bar Q^4 \cdot \bar \OO_J) (x_2) \rangle
\eeq
contributes to the type-B anomaly. Then, as one implements the supersymmetric Ward identity \eqref{eq:20} and starts moving the supercharges $Q$ around from the $x_1$-insertion in \eqref{cb2ae}, there are terms where the $Q$s land on the $x_2$ insertion and terms where the $Qs$ land on the $x$-insertion. Up to $x$-derivatives the latter terms vanish. Assuming once again that we can ignore the $x$-derivatives, we drop all terms where some $Q$s were moved on the $x$-insertion of the energy-momentum tensor. This suggests that we can recast the anomalous term of \eqref{cb2ae} as the anomalous term of \eqref{cb2ac}, up to a proportionality constant that coincides with the one in the unbroken phase \eqref{eq:2}, i.e.
\begin{align}
  \label{eq:26}
  \left\langle T(x) \mathcal{O}_{I}(x_1) \bar{\mathcal{O}}_{J}(x_2)\right\rangle \propto \Box_{x_2}^2\left\langle T(x) \Phi_{i}(x_2) \bar{\Phi}_{j}(x_2)\right\rangle \delta_I^i \delta_J^j\;.
\end{align}

To summarise, under the assumption that we can drop terms with $x$-derivatives, Poincar\'e supersymmetry guarantees that the anomalies of $\Phi_i \propto Q^4 \cdot \OO_I \delta{_i^I}$ and $ \OO_I$ are proportional to each other in all phases through a constant of proportionality which is independent of the exactly marginal couplings. Consequently, since $G_{i j}$ is covariantly constant in both the unbroken and broken phases, the same must be true for the $G_{IJ}$ anomaly of the $\Delta=2$ operators. Notice that the holomorphic part of the tangent bundle (which houses the holomorphic part of the exactly marginal deformations) is a product $\LL^4 \otimes \VV_2$ of four copies of the bundle of the left-moving supercharges $\LL$ and the bundle of $\Delta=2$ chiral primary operators $\VV_2$.\footnote{Analogous statements apply obviously to anti-holomorphic exactly marginal deformations, right-moving supercharges and $\Delta=2$ anti-chiral superconformal primaries.} Accordingly, the connection on the tangent bundle is a direct sum of the connection on $\LL^4$ and $\VV_2$, \cite{Papadodimas:2009eu,Baggio:2014ioa,Niarchos:2021iax}. However, on the anomalies $G_{ij}$ and $G_{IJ}$ only the part of the connection on $\VV_2$ contributes.

\subsection{Perturbative Checks}
\label{check}

As further evidence for the validity of the relation \eqref{eq:26} in phases with spontaneously broken conformal symmetry, we present an explicit test at leading order in perturbation theory on the Higgs branch of the 4D $\NN=2$ superconformal QCD (SCQCD) theory. We compute at tree-level the anomalies for $\Delta=2$ CBOs in the CFT and Higgs-branch phases $(G_2^{(\rm CFT)}, G_2^{(\rm Higgs)})$ and relate them to the anomalies of exactly marginal operators $(G_4^{(\rm CFT)}, G_4^{(\rm Higgs)})$ via the series of equalities
\beq
\label{checkaa}
G_2^{(\rm Higgs)} = G_2^{(\rm CFT)} = \frac{1}{192} \, G_4^{(\rm CFT)} = G_2^{(\rm CFT)}
~.
\eeq
The relation $G_2^{(\rm Higgs)} = G_2^{(\rm CFT)}$ is a special case of \eqref{eq:26}.

In 4D SCQCD there is a single $\Delta =2$ CBO $\OO$ and a single exactly marginal operator $\Phi$. In terms of the elementary fields that appear in the SCQCD Lagrangian (see e.g. \cite{Niarchos:2019onf} for a more detailed discussion on notation and conventions)
 \begin{align}
   \OO &= \Tr \varphi^2 \;,\cr
         \Phi &= 2 \, \Tr [\partial_\mu\varphi \partial^\mu\bar{\varphi} + i\lambda\sigma^\mu \partial_\mu \bar{\lambda}  +\frac14 F_{\mu \nu} F^{\mu \nu} + O(g) ] \, .
 \end{align}
These operators are related by supersymmetry as in \eqref{cb2aa}. In our conventions, the normalisation of the superalgebra is
\begin{equation}
\{Q^\II_\alpha,\bar Q_{\JJ\dot{\alpha}}\}=2\delta^\II_\JJ P_{\alpha\dot{\alpha}}
\, ,
\end{equation}
with $\alpha, \dot \alpha$ the 4D Lorentz spinor indices. We will perform a perturbative computation in SCQCD with arbitrary color group $G_C$. 

The broken-phase computations that we will present are performed in the Higgs-branch vacuum that was analysed in \cite{Niarchos:2019onf}, where
\begin{align}
 \langle Q^a_{Ii}\rangle&=v\delta_{I1}\delta^a_i
\label{HBvacuum}
\end{align}
with $v\in \mathbb R$. For $v\not=0$ the dilatation symmetry is spontaneously broken and a real massless dilaton $\sigma$ appears in the spectrum. This couples linearly to the energy-momentum tensor of the unbroken phase. By expanding the Lagrangian of $\mathcal N=2$ SCQCD around the vacuum \eqref{HBvacuum}, one can determine how the dilaton interacts with the elementary fields of the theory. In the following, we will be primarily interested in its couplings with the $A_\mu$ and $\varphi$ fields, which acquire a mass $m=gv$; these are
\begin{equation}
\begin{split}
\begin{tikzpicture}[scale=0.8, every node/.style={transform shape},baseline={([yshift=-.5ex]current bounding box.center)},node/.style={anchor=base,
    circle,fill=black!25,minimum size=18pt,inner sep=2pt}]]
\draw (0,0);
\draw[dashed,thick] (0,0) to (1.5,0);
\draw[thick] (1.5,0) to (2.7,1.2);
\draw[thick] (1.5,0) to (2.7,-1.2);
\draw (0.3,0.3) node {$\sigma$};
\draw (3,1.1) node {$\varphi^A$};
\draw (3,-1.1) node {$\bar\varphi^B$};
\end{tikzpicture} 
\end{split}\,\,=\,\,-i\frac{g^2v}{k}\delta^{AB}
\qquad \quad \qquad
\begin{split}
\begin{tikzpicture}[scale=0.8, every node/.style={transform shape},baseline={([yshift=-.5ex]current bounding box.center)},node/.style={anchor=base,
    circle,fill=black!25,minimum size=18pt,inner sep=2pt}]]
\draw (0,0);
\draw[dashed,thick] (0,0) to (1.5,0);
\draw[thick] (1.5,0) to (2.7,1.2);
\draw[thick] (1.5,0) to (2.7,-1.2);
\draw (0.3,0.3) node {$\sigma$};
\draw (3,1.1) node {$A_\mu^A$};
\draw (3.08,-1.1) node {$A^{\mu B}$};
\end{tikzpicture} 
\end{split}\,\,=\,\,-i\frac{g^2v}{2k}\delta^{AB}\quad.
\end{equation}
All computations will be performed directly in Euclidean space and the integrals will be evaluated using dimensional regularisation with ($\mu$ has the dimensions of a mass and $\epsilon>0$)
\begin{equation}
  \int \frac{d^{d}  l}{(2\pi)^{d }}\mapsto \mu^{2\epsilon}\int \frac{d^{2(2-\epsilon)}  l}{(2\pi)^{2(2-\epsilon)}}\quad. 
\end{equation}

\subsubsection{$\Delta = 2$ Anomaly in Conformal Phase}

The tree level 2-point function of $\OO$ in the CFT phase is obtained via simple Wick contraction of the scalar fields $\varphi$ (which can be carried out in two ways) \cite{Niarchos:2019onf}
\begin{equation}
\label{unbrokenDelta2}
\begin{split}
\langle \OO(p)\bar{\OO}(-p)\rangle\,\, = 
\begin{tikzpicture}[scale=0.8, every node/.style={transform shape},baseline={([yshift=-.5ex]current bounding box.center)},node/.style={anchor=base,
    circle,fill=black!25,minimum size=18pt,inner sep=2pt}]]
\draw (0,0);
\draw (0.8,0.5) node {$p$};
\draw[double,thick,->] (0,0) to (0.8,0);
\draw[double,thick] (0.8,0) to (1.5,0);
\draw[double,thick]  (3.8,0) to (3,0) ;
\draw (3.8,0.5) node {$p$};
\draw[double,thick,->] (3.8,0)  to  (4.1,0);
\draw[double,thick] (4.1,0)  to  (4.5,0);
\draw (0,-0.6) node {Tr$[\varphi^2]$};
\draw (4.8,-0.6) node {Tr$[\bar\varphi^2]$};
\draw (2.25,0.95) node {$\ell$};
\draw (2.25,-0.95) node {$\ell-p$};
\draw[thick] (1.5,0) .. controls(2.25,0.7).. (3,0) ;
\draw[thick] (1.5,0) .. controls(2.25,-0.7) .. (3,0);
\end{tikzpicture}
\end{split}
\hspace{+0.5cm}
=\quad
2\, \mathcal{C} \times\,I_1(p) \;.
\end{equation}
Here $\mathcal{C}$ is the color factor
\begin{equation}
\mathcal{C} = \Tr  [ T^A T^B ] \,  \Tr  [ T_A T_B ]
\end{equation}
with $A,B=1,\dots,\rank (G_c)$, while the integral $I_1(p)$ is the kinematic factor
\begin{equation}
I_1(p) = \int \frac{d^{d}  \ell}{(2\pi)^{d }} \frac{1}{\ell^2}\frac{1}{(\ell-p)^2} =  \frac{1}{(4\pi)^2}\left[\frac{1}{\epsilon}-\gamma +3 - \log\left(\frac{p_1^2}{4\pi\mu^2}\right)\right]\quad.
\end{equation}
According to \eqref{extractingCFT}, one then reads off
\begin{equation}
G_2^{(\rm CFT)}=2\frac{\mathcal{C}}{(2\pi)^4}\quad.
\end{equation}

\subsubsection{$\Delta = 4$ Anomaly in Conformal Phase}

At tree level, the 2-point function of the exactly marginal operators receives only two contributions:
\begin{equation}
\label{unbrokenDelta4}
\langle \Phi(p)\bar{\Phi}(-p)\rangle =  \,4 \times \,\,\,
\hspace*{-3.75cm}%
\begin{split}
\begin{tikzpicture}[scale=0.8, every node/.style={transform shape},baseline={([yshift=-.5ex]current bounding box.center)},node/.style={anchor=base,
    circle,fill=black!25,minimum size=18pt,inner sep=2pt}]]
\draw (0,0);
\draw (0.8,0.5) node {$p$};
\draw[double,thick,->] (0,0) to (0.8,0);
\draw[double,thick] (0.8,0) to (1.5,0);
\draw[double,thick]  (3.8,0) to (3,0) ;
\draw (3.8,0.5) node {$p$};
\draw[double,thick,->]  (4.5,0) to (3.6,0);
\draw (0.45,-0.5) node {Tr$[\partial_\mu\varphi \partial^\mu\bar{\varphi}]$};
\draw (4.1,-0.5) node {Tr$[\partial_\nu\varphi \partial^\nu\bar{\varphi}]$};
\draw (2.25,0.95) node {$\ell$};
\draw (2.25,-0.95) node {$\ell-p$};
\draw[thick] (1.5,0) .. controls(2.25,0.7).. (3,0) ;
\draw[thick] (1.5,0) .. controls(2.25,-0.7) .. (3,0);
\end{tikzpicture} 
\end{split}
+\,\,\,4\times
\hspace*{-0.2cm}%
\begin{split}
\begin{tikzpicture}[scale=0.8, every node/.style={transform shape},baseline={([yshift=-.5ex]current bounding box.center)},node/.style={anchor=base,
    circle,fill=black!25,minimum size=18pt,inner sep=2pt}] 
\draw (0,0);
\draw (0.8,0.5) node {$p$};
\draw[double,thick,->] (0,0) to (0.8,0);
\draw[double,thick] (0.8,0) to (1.5,0);
\draw[double,thick]  (3.8,0) to (3,0) ;
\draw (3.8,0.5) node {$p$};
\draw[double,thick,->]  (4.5,0) to (3.6,0);
\draw (0.2,-0.5) node {Tr$[\partial_{[\mu} A_{\nu]}\partial^{[\mu} A^{\nu]}]$};
\draw (4.55,-0.5) node {Tr$[\partial_{[\rho} A_{\sigma]}\partial^{[\rho} A^{\sigma]}]$};
\draw (2.25,0.95) node {$\ell$};
\draw (2.25,-0.95) node {$\ell-p$};
\draw[thick] (1.5,0) .. controls(2.25,0.7).. (3,0) ;
\draw[thick] (1.5,0) .. controls(2.25,-0.7) .. (3,0);
\end{tikzpicture}
\hspace{-3.8cm}
\end{split}
\quad \qquad\,\,
\end{equation}
The two individual diagrams
\begin{align}
\begin{split}
\begin{tikzpicture}[scale=0.8, every node/.style={transform shape},baseline={([yshift=-.5ex]current bounding box.center)},node/.style={anchor=base,
    circle,fill=black!25,minimum size=18pt,inner sep=2pt}]]
\draw (0,0);
\draw (0.8,0.5) node {$p$};
\draw[double,thick,->] (0,0) to (0.8,0);
\draw[double,thick] (0.8,0) to (1.5,0);
\draw[double,thick]  (3.8,0) to (3,0) ;
\draw (3.8,0.5) node {$p$};
\draw[double,thick,->]  (4.5,0) to (3.6,0);
\draw (0.45,-0.5) node {Tr$[\partial_\mu\varphi \partial^\mu\bar{\varphi}]$};
\draw (4.1,-0.5) node {Tr$[\partial_\nu\varphi \partial^\nu\bar{\varphi}]$};
\draw (2.25,0.95) node {$\ell$};
\draw (2.25,-0.95) node {$\ell-p$};
\draw[thick] (1.5,0) .. controls(2.25,0.7).. (3,0) ;
\draw[thick] (1.5,0) .. controls(2.25,-0.7) .. (3,0);
\end{tikzpicture} 
\end{split}\,\,\,\,\,\,\,\,\,\,\,=\,\,\mathcal{C}\times I_2(p)
\end{align}
\begin{align}
\begin{split}
\begin{tikzpicture}[scale=0.8, every node/.style={transform shape},baseline={([yshift=-.5ex]current bounding box.center)},node/.style={anchor=base,
    circle,fill=black!25,minimum size=18pt,inner sep=2pt}]]
\draw (0,0);
\draw (0.8,0.5) node {$p$};
\draw[double,thick,->] (0,0) to (0.8,0);
\draw[double,thick] (0.8,0) to (1.5,0);
\draw[double,thick]  (3.8,0) to (3,0) ;
\draw (3.8,0.5) node {$p$};
\draw[double,thick,->]  (4.5,0) to (3.6,0);
\draw (0.2,-0.5) node {Tr$[\partial_{[\mu} A_{\nu]}\partial^{[\mu} A^{\nu]}]$};
\draw (4.55,-0.5) node {Tr$[\partial_{[\rho} A_{\sigma]}\partial^{[\rho} A^{\sigma]}]$};
\draw (2.25,0.95) node {$\ell$};
\draw (2.25,-0.95) node {$\ell-p$};
\draw[thick] (1.5,0) .. controls(2.25,0.7).. (3,0) ;
\draw[thick] (1.5,0) .. controls(2.25,-0.7) .. (3,0);
\end{tikzpicture} 
\end{split}\,\,=\,\,2\,\mathcal{C}\times I_3(p)
\end{align}
are equally contributing Feynman processes, since the kinematic integrals $I_2(p)$, $I_3(p)$ are given by
\begin{align}
I_2(p)&=\int \frac{d^{d}  \ell}{(2\pi)^{d}} \frac{1}{\ell^2}\frac{1}{(\ell-p)^2}\times\left[\ell_\mu \ell^\nu(\ell-p)^\mu(\ell-p)_\nu\right]=\frac{p^4}{4}I_1(p)\\
I_3(p)&=\int \frac{d^{d} \ell}{(2\pi)^{d}} \frac{1}{l^2}\frac{1}{(\ell-p)^2}\times \ell_{[\mu}\ell^{[\rho}\delta^{\sigma]}_{\nu]}\times [(\ell-p)^{[\mu}(\ell-p)_{[\rho}\delta^{\nu]}_{\sigma]}]=\frac{p^4}{8}I_1(p)
\, .
\end{align}
Applying \eqref{extractingCFT} one extracts
\begin{equation}
\label{42CFT}
G_4^{(\rm CFT)}=192\, G_2^{(\rm CFT)}\, .
\end{equation}
The factor 192 is part of our conventions. This relation is an explicit tree-level check of the well-known general result \eqref{eq:2} \cite{Papadodimas:2009eu}.

\subsubsection{$\Delta = 2$ Anomaly in Higgs Phase}

Following \cite{Niarchos:2019onf}, we compute in the Higgs phase the 3-point function of $\OO,\bar{\OO}$ with the trace of the energy-momentum tensor $T=T^\mu\,_\mu$. At tree level, this 3-point function receives a contribution due to the dilaton field $\sigma$
\begin{equation}
\label{BrokenDelta2}
\begin{split}
\langle T(q) \OO(p_1)\bar{\OO}(p_2)\rangle\,\, = \,\,
\begin{tikzpicture}[scale=0.8, every node/.style={transform shape},baseline={([yshift=-.5ex]current bounding box.center)},node/.style={anchor=base,
    circle,fill=black!25,minimum size=18pt,inner sep=2pt}]]
\draw (5.8,0);
\draw[snake it,double,thick] (5.8,0) to (7.2,0);
\draw[dashed,thick,->] (7.2,0) to (7.95,0);
\draw[dashed,thick] (7.95,0) to (8.3,0);
\draw[thick,->] (8.3,0) to (8.9,0.6);
\draw[thick] (8.9,0.6) to (9.5,1.2);
\draw[thick,->]  (8.9,-0.6) to (8.7,-0.4) ;
\draw[thick]  (8.7,-0.4)  to (8.3,0) ;
\draw[thick]  (9.5,-1.2) to (8.9,-0.6);
\draw[thick,->] (9.5,1.2) to (9.5,0);
\draw[thick] (9.5,0) to (9.5,-1.2);
\draw[thick,double,->] (10.5,1.8) to (10,1.5);
\draw[thick,double] ((10,1.5) to (9.5,1.2);
\draw[thick,double,->] (10.5,-1.8) to  (10,-1.5);
\draw[thick,double] (10,-1.5) to  (9.5,-1.2);
\draw (5.9,0.5) node {$T$};
\draw (5.9,-0.5) node {$q$};
\draw (7.6,0.5) node {$\sigma$};
\draw (7.6,-0.5) node {$q$};
\draw (8.8,0.9) node {$\ell$};
\draw (8.5,-1) node {$-q+\ell$};
\draw (10.2,1.2) node {$p_1$};
\draw (10.2,-1.2) node {$p_2$};
\draw (10.65,2.3) node {Tr$[\varphi\varphi]$};
\draw (10.65,-2.3) node {Tr$[\bar\varphi\bar{\varphi}]$};
\draw (10.4,0) node {$\ell+p_1$};
\end{tikzpicture}
\end{split}
\,\,=\,\,
4\, \mathcal{C} \times\,I_4(q,p_1,p_2) \;.
\end{equation}
The combinatorial factor originates from the four possible Wick contractions between the $\varphi \bar{\varphi}$ coming out of the dilaton vertex and  the two operators $\Tr[\varphi\varphi]$, $\Tr[\bar\varphi\bar{\varphi}]$.
The kinematic integral $I_4(q,p_1,p_2)$ is given by 
\begin{align}
I_4(q,p_1,p_2) &=v^2g^2\int \frac{d^{d} \ell}{(2\pi)^{d}} \frac{1}{\ell^2+m^2}\frac{1}{(p_1+\ell)^2+m^2}\frac{1}{(\ell-q)^2+m^2}\overset{q\to0}{\longrightarrow}\frac{1}{2}\frac{1}{(4\pi)^2}
\, ,
\end{align}
where the mass in the broken phase is proportional to the Higgs vev $v$, $m^2=g^2v^2$.
From \eqref{extractingHB} one can read off the anomaly in the Higgs phase, as already discussed in \cite{Niarchos:2019onf}, which is
\begin{equation}
\label{2HCFT}
G_2^{(\rm Higgs)}=G_2^{(\rm CFT)}
\, .
\end{equation}

\subsubsection{$\Delta = 4$ Anomaly in Higgs Phase}

As in the conformal phase, in the Higgs phase the tree-level anomaly also arises from two equally contributing Feynman processes, with a $\varphi$ and $A^\mu$ field running respectively inside the loops, 
\begin{equation}
\label{unbrokenDelta4b}
\begin{split}
\langle T(q) \Phi(p_1)\bar{\Phi}(p_2)\rangle= 4 \times
\begin{tikzpicture}[scale=0.8, every node/.style={transform shape},baseline={([yshift=-.5ex]current bounding box.center)},node/.style={anchor=base,
    circle,fill=black!25,minimum size=18pt,inner sep=2pt}]]
\draw (5.8,0);
\draw[snake it,double,thick] (5.8,0) to (7.2,0);
\draw[dashed,thick,->] (7.2,0) to (7.95,0);
\draw[dashed,thick] (7.95,0) to (8.3,0);
\draw[thick,->] (8.3,0) to (8.9,0.6);
\draw[thick] (8.9,0.6) to (9.5,1.2);
\draw[thick,->]  (8.9,-0.6) to (8.7,-0.4) ;
\draw[thick]  (8.7,-0.4)  to (8.3,0) ;
\draw[thick]  (9.5,-1.2) to (8.9,-0.6);
\draw[thick,->] (9.5,1.2) to (9.5,0);
\draw[thick] (9.5,0) to (9.5,-1.2);
\draw[thick,double,->] (10.5,1.8) to (10,1.5);
\draw[thick,double] ((10,1.5) to (9.5,1.2);
\draw[thick,double,->] (10.5,-1.8) to  (10,-1.5);
\draw[thick,double] (10,-1.5) to  (9.5,-1.2);
\draw (5.9,0.5) node {$T$};
\draw (5.9,-0.5) node {$q$};
\draw (7.6,0.5) node {$\sigma$};
\draw (7.6,-0.5) node {$q$};
\draw (8.8,0.9) node {$\ell$};
\draw (8.5,-1) node {$-q+\ell$};
\draw (10.2,1.2) node {$p_1$};
\draw (10.2,-1.2) node {$p_2$};
\draw (10.3,2.2) node {Tr$[\partial_\mu\varphi\partial^\mu\bar{\varphi}]$};
\draw (10.3,-2.2) node {Tr$[\partial_\mu\varphi\partial^\mu\bar{\varphi}]$};
\draw (10.4,0) node {$\ell+p_1$};
\end{tikzpicture} 
\end{split}
+\,4\times
\begin{split}
\begin{tikzpicture}[scale=0.8, every node/.style={transform shape},baseline={([yshift=-.5ex]current bounding box.center)},node/.style={anchor=base,
    circle,fill=black!25,minimum size=18pt,inner sep=2pt}] 
\draw (5.8,0);
\draw[snake it,double,thick] (5.8,0) to (7.2,0);
\draw[dashed,thick,->] (7.2,0) to (7.95,0);
\draw[dashed,thick] (7.95,0) to (8.3,0);
\draw[thick,->] (8.3,0) to (8.9,0.6);
\draw[thick] (8.9,0.6) to (9.5,1.2);
\draw[thick,->]  (8.9,-0.6) to (8.7,-0.4) ;
\draw[thick]  (8.7,-0.4)  to (8.3,0) ;
\draw[thick]  (9.5,-1.2) to (8.9,-0.6);
\draw[thick,->] (9.5,1.2) to (9.5,0);
\draw[thick] (9.5,0) to (9.5,-1.2);
\draw[thick,double,->] (10.5,1.8) to (10,1.5);
\draw[thick,double] ((10,1.5) to (9.5,1.2);
\draw[thick,double,->] (10.5,-1.8) to  (10,-1.5);
\draw[thick,double] (10,-1.5) to  (9.5,-1.2);
\draw (5.9,0.5) node {$T$};
\draw (5.9,-0.5) node {$q$};
\draw (7.6,0.5) node {$\sigma$};
\draw (7.6,-0.5) node {$q$};
\draw (8.8,0.9) node {$\ell$};
\draw (8.5,-1) node {$-q+\ell$};
\draw (10.2,1.2) node {$p_1$};
\draw (10.2,-1.2) node {$p_2$};
\draw (10.2,2.2) node{Tr$[\partial_{[\mu}A_{\nu]}\partial^{[\mu}A^{\nu]}]$};
\draw (10.2,-2.2) node{Tr$[\partial_{[\rho}A_{\sigma]}\partial^{[\rho}A^{\sigma]}]$};
\draw (10.4,0) node {$\ell+p_1$};
\end{tikzpicture}
\end{split}
\end{equation}
The two Feynman diagrams above evaluate to
\begin{align}
\begin{split}
\begin{tikzpicture}[scale=0.8, every node/.style={transform shape},baseline={([yshift=-.5ex]current bounding box.center)},node/.style={anchor=base,
    circle,fill=black!25,minimum size=18pt,inner sep=2pt}]]
\draw (0,0);
\draw[snake it,double,thick] (5.8,0) to (7.2,0);
\draw[dashed,thick,->] (7.2,0) to (7.95,0);
\draw[dashed,thick] (7.95,0) to (8.3,0);
\draw[thick,->] (8.3,0) to (8.9,0.6);
\draw[thick] (8.9,0.6) to (9.5,1.2);
\draw[thick,->]  (8.9,-0.6) to (8.7,-0.4) ;
\draw[thick]  (8.7,-0.4)  to (8.3,0) ;
\draw[thick]  (9.5,-1.2) to (8.9,-0.6);
\draw[thick,->] (9.5,1.2) to (9.5,0);
\draw[thick] (9.5,0) to (9.5,-1.2);
\draw[thick,double,->] (10.5,1.8) to (10,1.5);
\draw[thick,double] ((10,1.5) to (9.5,1.2);
\draw[thick,double,->] (10.5,-1.8) to  (10,-1.5);
\draw[thick,double] (10,-1.5) to  (9.5,-1.2);
\draw (5.9,0.5) node {$T$};
\draw (5.9,-0.5) node {$q$};
\draw (7.6,0.5) node {$\sigma$};
\draw (7.6,-0.5) node {$q$};
\draw (8.8,0.9) node {$\ell$};
\draw (8.5,-1) node {$-q+\ell$};
\draw (10.2,1.2) node {$p_1$};
\draw (10.2,-1.2) node {$p_2$};
\draw (10.3,2.2) node {Tr$[\partial_\mu\varphi\partial^\mu\bar{\varphi}]$};
\draw (10.3,-2.2) node {Tr$[\partial_\mu\varphi\partial^\mu\bar{\varphi}]$};
\draw (10.4,0) node {$\ell+p_1$};
\end{tikzpicture} 
\end{split}\,\,\,\,\,\,\,\,\,\,\,=\,\,2\,\mathcal{C}\times I_5(q,p_1,p_2)
\, ,
\end{align}
\begin{align}
\begin{split}
\begin{tikzpicture}[scale=0.8, every node/.style={transform shape},baseline={([yshift=-.5ex]current bounding box.center)},node/.style={anchor=base,
    circle,fill=black!25,minimum size=18pt,inner sep=2pt}] 
\draw (0,0);
\draw[snake it,double,thick] (5.8,0) to (7.2,0);
\draw[dashed,thick,->] (7.2,0) to (7.95,0);
\draw[dashed,thick] (7.95,0) to (8.3,0);
\draw[thick,->] (8.3,0) to (8.9,0.6);
\draw[thick] (8.9,0.6) to (9.5,1.2);
\draw[thick,->]  (8.9,-0.6) to (8.7,-0.4) ;
\draw[thick]  (8.7,-0.4)  to (8.3,0) ;
\draw[thick]  (9.5,-1.2) to (8.9,-0.6);
\draw[thick,->] (9.5,1.2) to (9.5,0);
\draw[thick] (9.5,0) to (9.5,-1.2);
\draw[thick,double,->] (10.5,1.8) to (10,1.5);
\draw[thick,double] ((10,1.5) to (9.5,1.2);
\draw[thick,double,->] (10.5,-1.8) to  (10,-1.5);
\draw[thick,double] (10,-1.5) to  (9.5,-1.2);
\draw (5.9,0.5) node {$T$};
\draw (5.9,-0.5) node {$q$};
\draw (7.6,0.5) node {$\sigma$};
\draw (7.6,-0.5) node {$q$};
\draw (8.8,0.9) node {$\ell$};
\draw (8.5,-1) node {$-q+\ell$};
\draw (10.2,1.2) node {$p_1$};
\draw (10.2,-1.2) node {$p_2$};
\draw (10.2,2.2) node{Tr$[\partial_{[\mu}A_{\nu]}\partial^{[\mu}A^{\nu]}]$};
\draw (10.2,-2.2) node{Tr$[\partial_{[\rho}A_{\sigma]}\partial^{[\rho}A^{\sigma]}]$};
\draw (10.4,0) node {$\ell+p_1$};
\end{tikzpicture}
\end{split}\,\,\,\,\,\,\,\,=\,\,8\,\mathcal{C}\times I_6(q,p_1,p_2)
\, ,
\end{align}
with  the kinematical integrals $I_5(q,p_1,p_2)$ and $ I_6(q,p_1,p_2)$ given by
\begin{align}
I_5(q,p_1,p_2)&=v^2g^2\int \frac{d^{d} \ell}{(2\pi)^{d}} \frac{\ell_\mu}{\ell^2+m^2}\frac{(p_1+\ell)^\mu(p_1+\ell)^\nu}{(p_1+\ell)^2+m^2}\frac{(\ell-q)_\nu}{(\ell-q)^2+m^2}\cr
&\overset{q\to0}{\longrightarrow}\frac{1}{(4\pi)^2}\left(\frac{1}{8}p_1^4+...\right)
\, ,
\\
I_6(q,p_1,p_2)&=\frac12g^2v^2\int \frac{d^{d} \ell}{(2\pi)^{d}}\frac{\ell_{[\mu}\delta_{\nu]\alpha}}{\ell^2+m^2}\frac{(p_1+\ell)^{[\mu}\delta^{\nu]}_{[\sigma}(p_1+\ell)_{\rho]}}{(p_1+\ell)^2+m^2}\frac{(\ell-q)^{[\rho}\delta^{\sigma]\alpha}}{(\ell-q)^2+m^2}\cr
&\overset{q\to0}{\longrightarrow}\frac{1}{(4\pi)^2}\left(\frac{1}{32}p_1^4+...\right)
\, .
\label{FirstLoop}
\end{align}
Therefore, according to \eqref{extractingHB}, the anomaly for the marginal operators in the Higgs phase is given by
\begin{equation}\label{brcft}
G^{(\rm Higgs)}_4=G^{(\rm CFT)}_4
\, .
\end{equation}
Eqs.\ \eqref{2HCFT}, \eqref{42CFT}, \eqref{brcft} establish the announced sequence of relations in \eqref{checkaa}.

\section{Conclusions and Outlook}

In this paper we investigated the properties of type-B Weyl anomalies of integer-dimension operators on conformal manifolds. We 
presented evidence that such anomalies are covariantly constant on conformal manifolds in general phases of the theory, where conformal symmetry may be spontaneously broken. By explicitly constructing the corresponding anomaly functionals for operators of dimension $\Delta = 3,4,5$, and without relying on supersymmetry, we showed that $\nabla G= 0$ is a condition that guarantees WZ consistency. The anomaly functional for $\Delta =2$ operators was automatically WZ consistent, but we presented an independent argument in $\NN=2$ SCFTs using Poincar\'e supersymmetry that also implies $\nabla G=0$. This argument was explicitly checked to leading order in perturbation theory. It would be useful to examine if there is a more general, supersymmetry-independent argument, that proves $\nabla G=0$ for $\Delta=2$ anomalies.

One of the interesting features of the WZ-consistency analysis is that
it implies $\nabla G = 0$ in all phases of the theory, even when
conformal symmetry is spontaneously broken. We expect the WZ-consistency argument to hold for arbitrary-dimension integer operators, as a consequence of using integration by parts, but, as we showed, the cases of increasing scaling dimension involve increasingly complicated anomaly functionals where the curvature of the bundle of the integer-dimension operators plays a crucial role.

It is important to investigate further the stability of our WZ consistent anomaly functionals under possible deformations, e.g.\ under turning on nontrivial beta functions for sources/couplings. For instance, one such deformation can arise by having nontrivial beta functions for the exactly marginal couplings.\footnote{Using dimensional analysis, one can determine that potential beta functions for the sources $\lambda^I$ do not affect our construction of anomaly polynomials for $\Delta = 2,3,4,5$ to quadratic order in the $\lambda^I$.} New terms would then enter the computation of the WZ consistency condition through the anomaly functional for exactly marginal operators, \eqref{eq:30}. The simple option of having the standard beta function $\delta_{\sigma} \lambda^i\propto c^i_{JK} \lambda^J \lambda^K$, where $ c^i_{JK}$ is directly related to the 3-point function coefficient of $\langle \OO_i \OO_J \OO_K\rangle$, is not realised in our case, because the operators we consider have (by construction) fixed integer scaling dimensions along the conformal manifold and therefore vanishing coefficients $c^i_{JK} = 0$. However, we cannot rule out the existence of more general beta functions for marginal couplings that may receive contributions from the curvature on the vector bundle of operators. We hope to return to this point in the future.

\ack{We would like to thank M.~Broccoli and Z.~Komargodski for helpful discussions. E.A.\ is funded by the Royal Society Research Fellows Enhancement Award RGF/EA/180073. E.A.\ gratefully acknowledges support from the Simons Center for Geometry and Physics, Stony Brook University during a visit over the course of this work. C.P.\ is supported in part by a University Research Fellowship UF120032, and  in part through the STFC Consolidated Grant ST/P000754/1. The work of E.P.\ is partially supported by the GIF Research Grant I-1515-303./2019. }


\bibliography{anomalies}
\end{document}